\documentclass[journal=jcisd8,manuscript=article]{achemso}
\usepackage{graphicx} 
\usepackage{chemformula} 
\usepackage[T1]{fontenc} 
\usepackage{booktabs}
\usepackage[acronym]{glossaries}

\usepackage{caption}
\usepackage{subcaption}
\usepackage{rotating}

\usepackage[version=4]{mhchem} 
\usepackage{siunitx}
\DeclareSIUnit\atom{atom}
\DeclareSIUnit\step{step}
\DeclareSIUnit\angstrom{\text {Å}}

\usepackage{comment}
\usepackage{glossaries}

\usepackage[hidelinks]{hyperref}
\usepackage[capitalise,noabbrev]{cleveref}

\usepackage[inkscapeformat=png]{svg}

\usepackage{tabu}

\usepackage{xcolor}
\definecolor{mygreen}{RGB}{28,172,0}
\definecolor{mylilas}{RGB}{170,55,241}
\definecolor{codegreen}{rgb}{0,0.6,0}
\definecolor{codegray}{RGB}{105,105,105}
\definecolor{codepurple}{rgb}{0.58,0,0.82}
\definecolor{backcolour}{RGB}{240,240,240}

\DeclareMathOperator*{\argmaxA}{arg\,max}

\usepackage{listings}
\crefname{lstlisting}{listing}{listings}
\Crefname{lstlisting}{Listing}{Listings}
        
\newcommand{\code}[1]{\texttt{#1}}
\newcommand\YAMLkeystyle{\color{black}\ttfamily\fontsize{12}{12}} 
\newcommand\YAMLvaluestyle{\color{blue}\mdseries}

\makeatletter

\newcommand\language@yaml{yaml}

\expandafter\expandafter\expandafter\lstdefinelanguage
\expandafter{\language@yaml}
{
  keywords={true,false,null,y,n},
  keywordstyle=\color{darkgray}\bfseries,
  basicstyle=\linespread{1.1}\YAMLkeystyle,
  sensitive=false,
  comment=[l]{\#},
  morecomment=[s]{/*}{*/},
  commentstyle=\color{codegreen}\ttfamily,
  stringstyle=\YAMLvaluestyle\ttfamily,
  moredelim=[l][\color{orange}]{\&},
  moredelim=[l][\color{magenta}]{*},
  moredelim=**[il][\YAMLkeystyle{:}\YAMLvaluestyle]{:},   % switch to value style at :
  morestring=[b]',
  morestring=[b]",
  breaklines=true,
  literate =    {---}{{\ProcessThreeDashes}}3
                {>>>}{{\textcolor{codegreen}\textgreater\textcolor{codegreen}\textgreater\textcolor{codegreen}\textgreater}}3
                {|}{{\textcolor{codegreen}\textbar}}1 
                {\ -\ }{{\mdseries\ -\ }}3,
}

\lst@AddToHook{EveryLine}{\ifx\lst@language\language@yaml\YAMLkeystyle\fi}
\makeatother

\newcommand\ProcessThreeDashes{\llap{\color{cyan}\mdseries-{-}-}}

\SectionNumbersOn
\author{Moritz R. Sch\"{a}fer}
\affiliation[TheoChem]{Institute for Theoretical Chemistry, University of Stuttgart, Pfaffenwaldring 55, 70569 Stuttgart, Germany}
\altaffiliation{contributed equally}

\author{Nico Segreto}
\affiliation[TheoChem]{Institute for Theoretical Chemistry, University of Stuttgart, Pfaffenwaldring 55, 70569 Stuttgart, Germany}
\altaffiliation{contributed equally}

\author{Fabian Zills}
\affiliation[ICP]{Institute for Computational Physics, University of Stuttgart, Allmandring 3, 70569 Stuttgart, Germany}
\author{Christian Holm}
\affiliation[ICP]{Institute for Computational Physics, University of Stuttgart, Allmandring 3, 70569 Stuttgart, Germany}

\author{Johannes K\"{a}stner}
\affiliation[TheoChem]{Institute for Theoretical Chemistry, University of Stuttgart, Pfaffenwaldring 55, 70569 Stuttgart, Germany}
\email{kaestner@theochem.uni-stuttgart.de}

\title[Apax]
  {Apax: A Flexible and Performant Framework For The Development of Machine-Learned Interatomic Potentials}

\newacronym{rtil}{RTIL}{room-temperature ionic liquid}
\newacronym{mlp}{MLP}{machine-learned potential}
\newacronym{mlip}{MLIP}{machine-learned interatomic potential}
\newacronym{dft}{DFT}{density functional theory}
\newacronym{md}{MD}{molecular dynamics}
\newacronym{rdf}{RDF}{radial distribution function}
\newacronym{aimd}{AIMD}{\textit{ab initio} molecular dynamics}
\newacronym{mae}{MAE}{mean absolute error}
\newacronym{mse}{MSE}{mean squared error}
\newacronym{rmse}{RMSE}{root mean squared error}
\newacronym{nll}{NLL}{negative loglikelihood}
\newacronym{msd}{MSD}{mean squared displacement}
\newacronym{ase}{ASE}{Atomic Simulation Environment}
\newacronym{gmnn}{GMNN}{Gaussian Moment Neural Network}
\newacronym{apax}{\texttt{apax}}{Atomistic learned potentials in JAX}
\newacronym{jit}{JIT}{just-in-time}
\newacronym{xla}{XLA}{Accelerated Linear Algebra}
\newacronym{bal}{BAL}{batch active learning}
\newacronym{pbe}{PBE}{Perdew-Burke-Ernzerhof}
\newacronym{gth}{GTH}{Goedecker-Teter-Hutter}
\newacronym{bmim}{\ce{BMIM+BF4-}}{1-Butyl-3-methylimidazolium tetrafluoroborate}
\newacronym{emim}{\ce{EMIM+BF4-}}{1-ethyl-3-methylimidazolium tetrafluoroborate}
\newacronym{pes}{PES}{Potential Energy Surface}
\newacronym{tl}{TL}{Transfer Learning}
\newacronym{udd}{UDD}{uncertainty-driven dynamics}
\newacronym{iae}{IAE}{integrated absolute errors}
\newacronym{crps}{CPRS}{continuous ranked probability score}

\begin{document}
\begin{tocentry}
    \includegraphics[width=\linewidth]{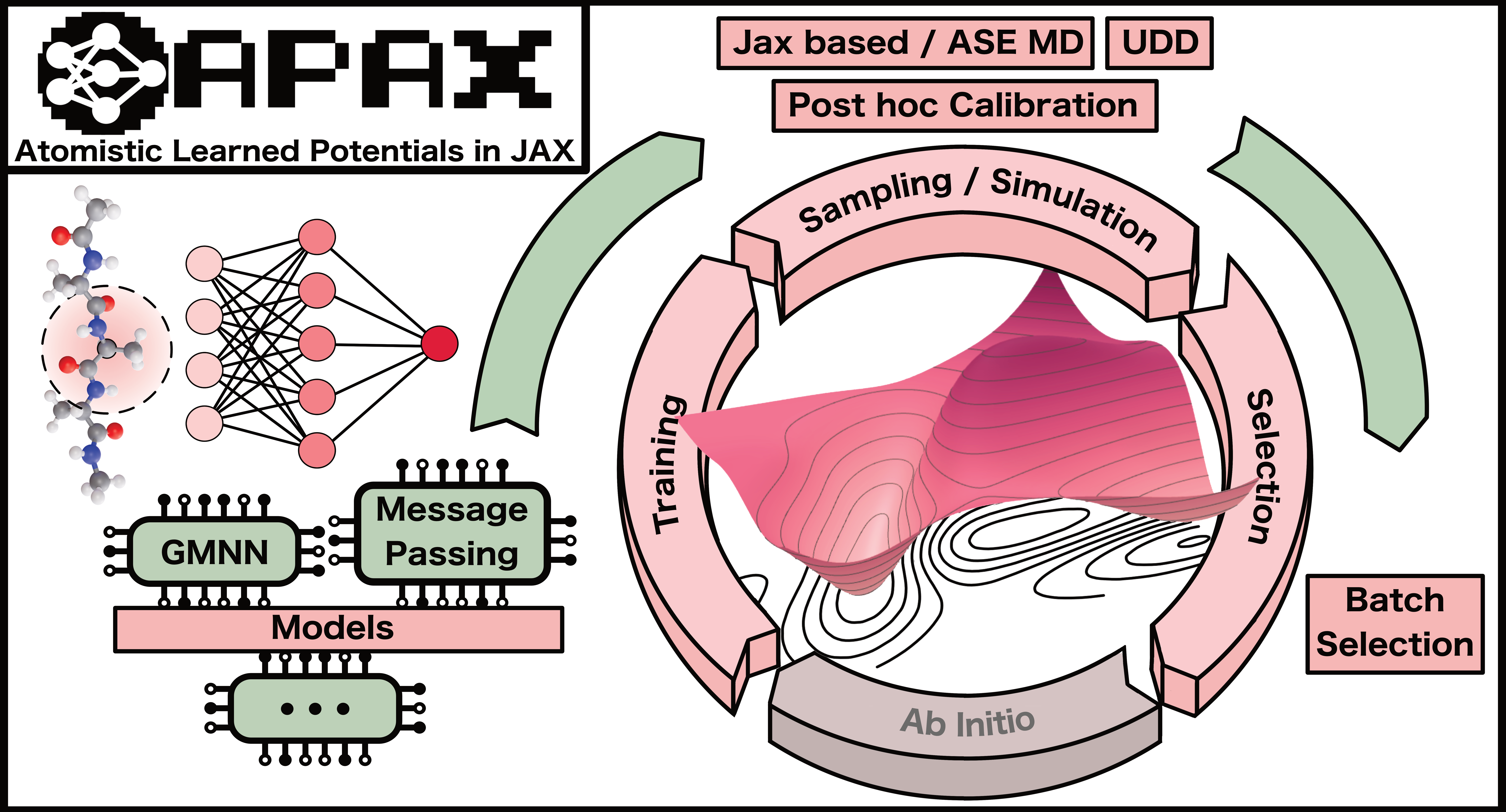}
\end{tocentry}

\begin{abstract}
    We introduce \gls{apax}, a flexible and efficient open source software package for training and inference of machine-learned interatomic potentials.
    Built on the JAX framework, \gls{apax} supports GPU acceleration and implements flexible model abstractions for fast development.
    With features such as kernel-based data selection, well-calibrated uncertainty estimation, and enhanced sampling, it is tailored to active learning applications and ease of use.
    The features and design decisions made in \gls{apax} are discussed before demonstrating some of its capabilities.
    First, a data set for the room-temperature ionic liquid \ce{EMIM+BF4-} is created using active learning. It is highlighted how continuously learning models between iterations can reduce training times up to 85~\% with only a minor reduction of the models' accuracy.
    Second, we show good scalability in a data-parallel training setting.
    We report that a Gaussian Moment Neural Network model, as implemented in \gls{apax}, achieves higher accuracy and up to 10 times faster inference times than a performance-optimized Allegro model.
    A recently published Li$_3$PO$_4$ dataset, reported with comparable accuracy and inference performance metrics, is used as a point of comparison.
    Moreover, the inference speeds of the available simulation engines are compared.
    Finally, to highlight the modularity of \gls{apax}, an equivariant message-passing model is trained as a shallow ensemble and used to perform uncertainty-driven dynamics.

\end{abstract}

\section{Introduction}

Traditionally, computational methods to describe the interactions of atoms in chemical systems are grouped in quantum mechanical methods, like \gls{dft}, and empirical force fields.
While the former group of methods can achieve great accuracy, it is many orders of magnitude slower than much less accurate empirical force fields.
The emergence of \glspl{mlip} has led to a paradigm shift in the field of molecular simulation\cite{unkeMachineLearningForce2021,behlerMachineLearningPotentials2021}.
Within only a few years, the scope of these potentials has progressed from simple systems requiring thousands of training data points for adequate accuracy\cite{artrithHighdimensionalNeuralNetwork2012,schuttSchNetDeepLearning2018a} to modeling complex systems of tens of thousands of atoms using ever smaller amounts of data\cite{musaelianLearningLocalEquivariant2023,batatiaMACEHigherOrder2022, vandermauseFlyActiveLearning2020}.
Trained to reproduce the energies and forces of quantum mechanical calculations, \glspl{mlip} can closely match the accuracy of the reference methods while achieving linear scaling with the number of particles.

Nowadays, plenty of \gls{mlip} codes are publicly available.
However, most of them serve primarily as a reference implementation of a particular \gls{mlip} model architecture \cite{batznerE3equivariantGraphNeural2022, batatiaMACEHigherOrder2022,unkePhysNetNeuralNetwork2019} 
and do not aim to be simultaneously modular and flexible method development frameworks, as well as performant training and inference engines.
Among these frameworks are SchnetPack\cite{schuttSchNetDeepLearning2018a}, DeepMDkit\cite{wangDeePMDkitDeepLearning2018,zengDeePMDkitV2Software2023},  FeNNol\cite{pleFeNNolEfficientFlexible2024} and chemtrain\cite{fuchsChemtrainLearningDeep2024}.
Schnetpack is designed for highly configurable model and training workflows and comes with a LAMMPS\cite{thompsonLAMMPSFlexibleSimulation2022} pair style for all its models.
DeepMD features the various DeepMD models\cite{NEURIPS2018_e2ad76f2,zengDeePMDkitV2Software2023} and supports distributed \gls{md}, allowing it to scale to very large systems\cite{zhangDeepPotentialMolecular2018}.
The FeNNol package emphasizes building hybrid ML/MM models\cite{pleForcefieldenhancedNeuralNetwork2023} as well as an integration into the TinkerHP\cite{lagardereTinkerHPMassivelyParallel2018} ecosystem.
Lastly, chemtrain focuses on advanced training setups, such as coarse-graining and differential trajectory reweighting\cite{camposDifferentiableNeuralNetworkForce2022a}.
The presented work introduces \gls{apax}\cite{schaferApaxhubApaxV0702024}, a Python package for training and applying machine-learned potentials based on the JAX\cite{jax2018github} framework.
In contrast to the packages discussed above, the primary methodological focus of \gls{apax} is the iterative generation of datasets, often referred to as learning on the fly or active learning.
From a software standpoint, it emphasizes performance and flexibility.

Active learning runs are typically initialized with only a handful of training data points.
In an iterative manner, preliminary models perform sampling \gls{md} simulations that serve as a pool of candidate configurations.
Models are retrained after recomputing the most informative snapshots with a quantum chemistry method.
The cycle can be repeated until sufficient model accuracy is achieved\cite{cohnNeuralNetworkExploration1996,liMolecularDynamicsOntheFly2015,podryabinkinActiveLearningLinearly2017,smithLessMoreSampling2018,shuaibiEnablingRobustOffline2020,zaverkinExplorationTransferableUniformly2021a}. 

Typically, transfer learning is used to adapt a pre-trained model to a related dataset, leveraging prior knowledge to improve performance and accelerate convergence on new data\cite{panSurveyTransferLearning2010a,smithApproachingCoupledCluster2019a,smithANI1ccxANI1xData2020a,zaverkinTransferLearningChemically2023}.
The approach can also be applied to active learning: rather than optimizing all parameters from scratch, parameters from the previous iteration are further refined.
Such continuous learning strategies can reduce the necessary number of epochs, thereby lowering overall training costs.

Active learning workflows have various requirements for the methods employed at every step.
Utilizing models that can be trained quickly is advantageous, as active learning involves numerous retraining cycles.
The fast simulation times of the \gls{gmnn}\cite{zaverkinGaussianMomentsPhysically2020,zaverkinFastSampleEfficientInteratomic2021} model, coupled with its good predictive accuracy, make it an ideal model for active learning workflows and large-scale simulations.
The \gls{gmnn} has been successfully applied to a variety of systems from ionic liquids\cite{zillsMachineLearningdrivenInvestigation2024} to astrochemical surfaces\cite{mol20a,mol21a} and other systems\cite{gubaevPerformanceTwoComplementary2023,zaverkinThermallyAveragedMagnetic2022}.
Nonetheless, \gls{apax} offers modular model abstractions that are compatible with any atom-centered representation.
During sampling simulations, it is necessary to estimate the model's uncertainty and stop trajectories from entering unphysical regimes\cite{zillsCollaborationMachineLearnedPotentials2024}.
The shallow ensembles recently proposed by \citeauthor{kellnerUncertaintyQuantificationDirect2024a}\cite{kellnerUncertaintyQuantificationDirect2024a}, trained on probabilistic loss functions, offer a reliable way to estimate uncertainties.
Additionally, biasing the dynamics towards regions of higher uncertainty can enhance the sampling during those simulations, increasing the informativeness of encountered configurations\cite{kulichenkoUncertaintydrivenDynamicsActive2023,vanderoordHyperactiveLearningDatadriven2023}.
A new set of data points needs to be selected once a sampling simulation is completed.
These should be, at the same time, informative and sufficiently different from each other.
Therefore, \gls{apax} implements the batch data selection methods\cite{zaverkinExploringChemicalConformational2022} developed in the Kästner group.

The study is structured as follows.
We begin by introducing \glspl{mlip}, particularly \gls{gmnn} and the other theoretical methods used in the \gls{apax} package.
Afterward, we begin the demonstrations by creating a dataset for the \gls{emim} \gls{rtil} using active learning.
Here, we demonstrate how reusing model parameters from the previous iteration can be used to reduce the training time of the current iteration.
Next, we analyze the training performance for the \gls{gmnn} and equivariant message passing models in a data-parallel setting.
The inference performance is investigated for the \gls{ase} and the internal \gls{md} engine\cite{schoenholzJAXMDFramework2021} for different ensemble kinds and sizes.
The performance studies use the \gls{emim} dataset constructed during active learning and subsequently the \ce{Li3PO4} dataset by \citeauthor{batznerE3equivariantGraphNeural2022}\cite{batznerE3equivariantGraphNeural2022} where inference speeds are available for the Allegro\cite{musaelianLearningLocalEquivariant2023} and SO3krates\cite{frankEuclideanTransformerFast2024} models.
Finally, we demonstrate the modularity of \gls{apax}'s design by training an equivariant message-passing model as a shallow ensemble and investigate the sampling advantages offered by \gls{udd}\cite{kulichenkoUncertaintydrivenDynamicsActive2023}.

\section{Methods}

\subsection{Machine Learning Interatomic Potentials}
Given an atomic configuration $S$ consisting of coordinates $\boldsymbol{R}$ and atomic numbers $Z$, potentials used in \gls{md} map from $S$ to a potential energy $E$.
For most \glspl{mlip}, the locality of atomic interactions is assumed.
As a result, additive functional forms are used, with atomic contributions dependent on their respective local environment\cite{behlerGeneralizedNeuralNetworkRepresentation2007a,unkeMachineLearningForce2021}:

\begin{align}\label{eq:energy_sum}
    E(S, \boldsymbol{\theta}) = \sum_i^{N_{\text{atoms}}} E_i( \boldsymbol{G}_i, \boldsymbol{\theta})
\end{align}

Here, $\theta$ is the set of parameters that are adjusted during training and $\boldsymbol{G}_i$ a representation of the local atomic environment around atom $i$.
Individual \glspl{mlip} differ in the particular representation of local environments and the regression model used to predict atomic energies.
One particular \gls{mlip}, developed in the Kästner group, is the \gls{gmnn} model \cite{zaverkinGaussianMomentsPhysically2020,zaverkinFastSampleEfficientInteratomic2021}.
The model is composed of the Gaussian Moment descriptor and feed-forward neural networks to compute atomic energies.
The descriptor maps the pairwise distance vectors between each atom and its local neighbors to a feature vector which is invariant under translations and rotations of the system.
A smooth radial neighborhood density is constructed from a set of basis functions. 
Many possible basis functions have been proposed in the literature, but here we implement equidistant Gaussians\cite{schuttSchNetDeepLearning2018a} and the non-orthogonal Bessel-like functions of \citeauthor{kocerNovelApproachDescribe2019}\cite{kocerNovelApproachDescribe2019}
These are linearly combined by element-pair coefficients $\beta_{ij}$ to form a radial basis $R$.
Angular information is captured by cartesian moments constructed with the distance vectors up to some rotation order $L$.
Pairwise contributions are summed over all neighboring atoms $j$.

\begin{align}\label{eq:basis_fn}
    \Psi_{i, L, s} = \sum_{j \neq i} R_{Z_i, Z_j, s}(r_{ij}, \beta_{ij}) \boldsymbol{\hat{r}}_{ij}^{\otimes L}
\end{align}

The final descriptor is obtained from a set of full tensor contractions, which are implemented up to $L=3$ and 4-body terms.

\begin{align}\label{eq:contraction}
    G_{i, s_1, s_2} &= (\Psi_{i, 1, s_1})_a (\Psi_{i, 1, s_2})_a  \nonumber \\ 
    &\;\; \vdots \\
    G_{i, s_1, s_2, s_3} &= (\Psi_{i, 1, s_1})_a (\Psi_{i, 3, s_2})_{a,b,c} (\Psi_{i, 2, s_3})_{b,c} \nonumber
\end{align}

Atomic energies are predicted from neural networks as $E_i = \mathrm{NN}(\boldsymbol{G}_i)$ and scaled and shifted by per-element parameters $\sigma_{Z_i}$ and $\mu_{Z_i}$ before being summed up according to \cref{eq:energy_sum}.

\begin{align}\label{eq:scale_shift}
    E_i = \sigma_{Z_i} \cdot \text{NN}(\boldsymbol{G}_i) + \mu_{Z_i}
\end{align}

In addition to the \gls{gmnn} model, we also implement an equivariant message passing model, EquivMP, similar to NequIP\cite{batznerE3equivariantGraphNeural2022}.

\subsection{Uncertainty Quantification}
During the sampling simulations of early active learning iterations, it is likely that the model is not yet accurate enough to simulate a stable \gls{md} trajectory.
It is thus necessary to use stopping criteria for the simulation, which terminate the run before the system explores unphysical configurations.
The true deviation from the reference method is not known during a sampling simulation. Therefore, the model's uncertainty, as a natural choice for such an error estimate, is used as a stopping criterion.

Ensembling multiple independently trained models has proven to be a straightforward way to estimate uncertainties\cite{seungQueryCommittee1992,artrithHighdimensionalNeuralNetwork2012,smithLessMoreSampling2018}.
Given the predictions of $N_{\text{ens}}$ members, the uncertainty of the ensemble can be obtained from the sample standard deviation $\sigma_x$ of the predictions\cite{apetersonAddressingUncertaintyAtomistic2017}.
For conciseness, we use $x \in \{E, F_i\}$ as the equations below can be used for both total energies $E$ and atomic forces $F_i$ with an implied sum over components and atoms for the latter.

\begin{align}
    \sigma_x  & = \sqrt{ \frac{1}{N_{\text{ens}}} \sum_m^{N_{\text{ens}}} (x^{(m)} - \bar{x})^2 } \label{eq:uncertainty}
\end{align}

Compared to other uncertainty quantification methods\cite{xieUncertaintyawareMolecularDynamics2023, zhuFastUncertaintyEstimates2023}, model ensembles typically achieve lower validation errors than a single model\cite{carreteDeepEnsemblesVs2023a}.
However, model ensembles face two challenges.
First, the training and inference time increase linearly with the number of ensemble members.
Second, the estimated uncertainties are usually miscalibrated\cite{guo17aCalibration, kuleshov18a_ensemble}, i.e., there is a low correlation between estimated uncertainty and actual prediction error.

Instead of ensembling full models, the recently proposed shallow ensembles\cite{kellnerUncertaintyQuantificationDirect2024a} obtain multiple model predictions by only ensembling the last neural network layer.
As a result, energy uncertainties can be computed at negligible additional cost.
The forces can be evaluated in two equivalent ways, either as the gradient of the mean energy or the mean of the Jacobian of the ensemble energy predictions.

\begin{align}
    F_i = \frac{\partial \bar{E}}{\partial r_i} = \frac{1}{N_{\text{ens}}} \sum_m^{N_{\text{ens}}} F_i^{(m)}
    \label{eq:jacobian}
\end{align}

While the latter is more computationally expensive, it allows computing force uncertainties via \cref{eq:uncertainty} and is still cheaper than the equivalent full ensemble.
Further, these shallow ensembles are trained on probabilistic loss functions, like the \gls{nll}, or \gls{crps}\cite{Gneiting01032007}, instead of the typical homoscedastic loss functions such as \gls{mse} and Huber\cite{huberRobustEstimationLocation1964}.
Using probabilistic losses ensures that both the means and standard deviations of ensemble predictions follow the empirical distribution.

\begin{align}
    \mathrm{NLL} &= \frac{1}{2} \left[ \frac{(x - x_\text{ref})^2}{\sigma_x^2} + \log 2 \pi \sigma_x^2 \right] \\ 
    \text{CRPS}_{\mathcal{N}}(|x - x_\text{ref}|, \sigma) &= \sigma \left\{ \frac{|x - x_\text{ref}|}{\sigma} \left[ 2\Phi\left( \frac{|x - x_\text{ref}|}{\sigma} \right) - 1 \right]
+ 2\varphi\left( \frac{|x - x_\text{ref}|}{\sigma} \right) - \frac{1}{\sqrt{\pi}} \right\}
    \label{eq:nll}
\end{align}

Additionally, it is possible to correct a model's calibration \textit{post hoc} by scaling the predicted uncertainties by a (sample-dependent) factor determined from a validation set.
We refer to the discussion by \citeauthor{kellnerUncertaintyQuantificationDirect2024a}\cite{kellnerUncertaintyQuantificationDirect2024a} and \citeauthor{rahimiPosthocCalibrationNeural2022}\cite{rahimiPosthocCalibrationNeural2022} for more details on post hoc calibration.

\subsection{Molecular Dynamics and Biased Sampling}
In atomistic simulations, computing thermodynamic observables requires averaging over configurations sampled from the system's Boltzmann distribution\cite{mccammonDynamicsFoldedProteins1977,karplusMolecularDynamicsSimulations1990}.
Such configurations are typically obtained through \gls{md} simulations, which involve the integration of the equations of motion of the atoms in the system.
In addition to production simulations, preliminary \glspl{mlip} can be used to sample candidate configurations more efficiently than with \textit{ab initio} methods. 
However, long \gls{md}-trajectories are necessary to sample slow degrees of freedom and rare events\cite{beveridgeFreeEnergyMolecular1989,kollmanFreeEnergyCalculations1993,laioMetadynamicsMethodSimulate2008,kastnerUmbrellaSampling2011}.
So-called enhanced sampling methods can drastically reduce the simulation time required to explore a system's free energy surface, often by adding a bias $E_{\text{b}}$ to the potential energy.
To generate candidate configurations, several enhanced sampling methods have emerged in recent years\cite{yooMetadynamicsSamplingAtomic2021a,vanderoordHyperactiveLearningDatadriven2023}.
One such method is \gls{udd}\cite{kulichenkoUncertaintydrivenDynamicsActive2023}, which uses the predicted uncertainty of \gls{mlip} ensembles to drive the dynamics of a system towards regions of high uncertainty.

\begin{align}
    E_{\text{b}}(\sigma_{\text{E}}^2) = A \left[ \exp \left( - \frac{\sigma_{\text{E}}^2}{N_{\text{ens}} N_{\text{atoms}} B^2} \right) - 1   \right]
    \label{eq:udd}
\end{align}

The bias potential includes two parameters $A$ and $B$, which denote the maximal strength of the bias potential and its width, respectively.

\subsection{Batch Active Learning}
Once a sampling trajectory is completed, new data points can be selected to expand the dataset.
Here, \cref{eq:uncertainty} could be used as well to choose those configurations with the largest predicted uncertainty.
However, doing so in a greedy manner is likely to select adjacent samples from the trajectory, limiting the diversity of selected configurations.
\Gls{bal} methods allow incorporation of the similarity of new data points to already chosen ones in the selection process, introducing more diversity into the selection.
More concretely, given a pool of data $\mathcal{D}_\text{pool} = \{S_1, ... , S_n\}$, the \gls{bal} task consists in finding a batch $\mathcal{D}_\text{batch} = \{S^*_1, ... , S^*_b\} \subset \mathcal{D}_\text{pool}$ which maximizes an acquisition function $a$, dependent on the model parameters\cite{kirschBatchBALDEfficientDiverse2019a}.

\begin{align}\label{eq:bal}
    \mathcal{D}_\text{batch} = \argmaxA_{\{S_1, ... , S_b\} \subset \mathcal{D}_\text{pool}} a(\{S_1, ... , S_b\}, \boldsymbol{\theta})
\end{align}

The acquisition function is typically constructed in such a way as to ensure diversity between selected data points or other criteria of the underlying data distribution\cite{zaverkinExploringChemicalConformational2022}.
In the present work, we use a greedy maximum distance selection with a last-layer gradient feature map $\phi_{\text{ll}}$.

\begin{align}
    \phi_{\text{ll}}(S) &= \nabla_{\theta_{\text{ll}}} E(S, \theta) \label{eq:lastlayer}\\
    S &= \argmaxA_{S \in \mathcal{D}_\text{pool} / \mathcal{D}_\text{batch}} \min_{S' \in \mathcal{D}_\text{pool} \cup \mathcal{D}_\text{batch}} \Delta(S, S') \label{eq:maxdist}
\end{align}

Here, $\Delta(S, S') = || \phi(S) - \phi(S') ||_2 $ is the distance between feature vectors.
\Cref{eq:maxdist} is applied iteratively, meaning that each application yields the next most distant structure in feature space.
For further details, we refer to the original publication\cite{zaverkinExploringChemicalConformational2022}.

\section{Software Architecture}

The \gls{apax} package is based on the JAX framework\cite{jax2018github}, allowing for GPU acceleration and \gls{jit} compilation.
In contrast to other machine learning frameworks, such as PyTorch\cite{paszke19a-pytorch} and TensorFlow\cite{tensorflow2015-whitepaper}, JAX is based on principles from functional programming, such as pure functions and composable function transformations.
Its functional purity and static computation graphs allow the \gls{xla} compiler to emit highly optimized code, which usually outperforms other frameworks\cite{frankEuclideanTransformerFast2024}.
The function transformations enable users to quickly build up complex functions from simpler ones, e.g., transforming a function that returns the potential energy of an atomistic system into one that returns the energy and corresponding forces.
\gls{apax} embraces the philosophy of JAX and implements a lot of its functionality in terms of such transformations.

\begin{figure}
    \centering
    \includegraphics[width=1.0\textwidth]{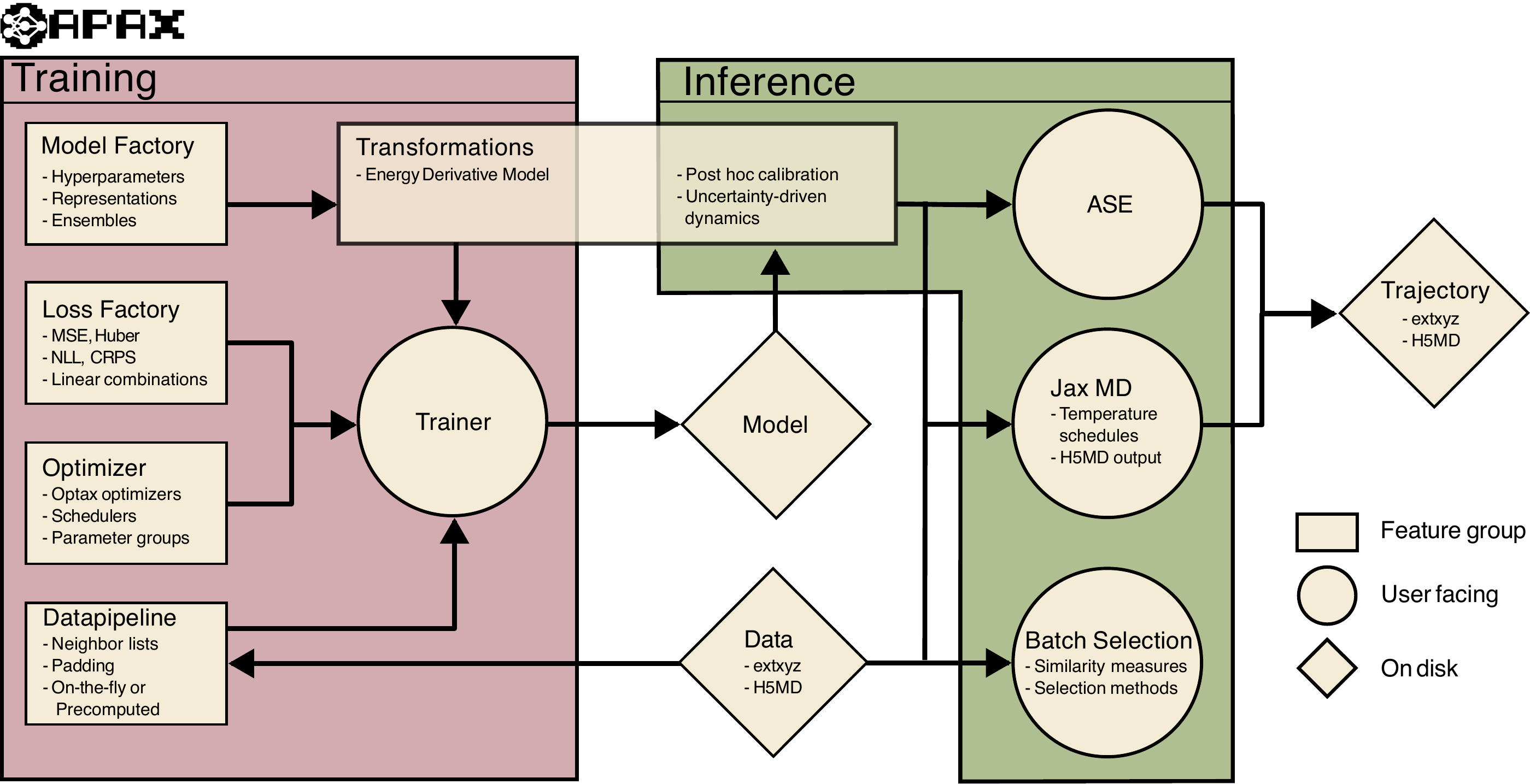}
    \caption{Overview of the features and code structure of the \gls{apax} package.
    Circles represent user-facing functionalities, rectangles internal feature groups, and diamond shapes data stored on disk.
    }
    \label{fig:software}
\end{figure}

In the remainder of the section, we will discuss the structure of \gls{apax} in more detail.
A high-level overview of the data flow and feature availability in the package is displayed in \cref{fig:software}.
All features of \gls{apax} are exposed to the IPSuite\cite{zillsCollaborationMachineLearnedPotentials2024} workflow manager, co-developed by the authors.
It separates the model and deep-learning framework-specific code from general-purpose functionality, such as workflow construction, data versioning, and sharing and model evaluation methods.

\subsection{Command-Line Interface and Configuration}
\gls{apax} provides a command-line interface to utilize its training and JAX-based \gls{md} functionalities.
To make the initial usage experience as streamlined as possible, the commands \verb|apax template train/md| can be used to create templates for training and \gls{md} configuration files.
As a file format, we have chosen YAML for its easy human and machine readability.
Further, the Pydantic library is used to validate inputs, allowing users to locally check for spelling mistakes of keywords and the correctness of supplied data types before submitting jobs to queuing systems on high performance computers.
Some examples of errors that can be caught are given in \Cref{lst:pydantic_val_error}.
In the remainder of the section, we illustrate parts of the configuration files where appropriate.

\subsection{Data Pipeline}
Training \glspl{mlip} requires iterating through a collection of training structures hundreds or thousands of times while passing this data to the model for evaluation.
The loading of atomistic data requires some careful considerations, especially in a JAX-based framework.
First, the training dataset needs to be in an appropriate format.
\gls{apax} supports reading atomistic structures into \gls{ase} Atoms objects from any \gls{ase} readable format, as well as H5MD\cite{debuylH5MDStructuredEfficient2014} files \textit{via} the ZnH5MD\cite{zillsZincwareZnH5MDV0362024} library.
The extended XYZ from \gls{ase} and H5MD are the most suitable file formats due to their flexibility and, in the latter case, IO performance.
As a result, many existing literature datasets can be read directly.
Secondly, one needs to account for the static computation graphs required by JAX in the preprocessing of data:
All input shapes, e.g., the number of atoms and neighbors, need to be known at compile time, with deviations causing recompilation.
While a single recompilation takes only a few seconds, recompiling for every batch in a dataset would be impractically slow.
There are two options for handling differently sized systems implemented in \gls{apax}.

For the first option, referred to as ``cached'', the systems with the largest number of atoms and neighbors are identified.
All samples in the dataset are then padded with zeros to the largest number of atoms, for per-atom quantities such as positions, and the largest number of neighbors, for the neighbor list, respectively.
As a result, the training step function only needs to be compiled once, and the additional compute used for the padding is negligible when all structures in the dataset have the same or similar sizes.
Instead of padding every sample to the same size, it is also possible to bin system sizes.
By allowing for a few recompilations, it is possible to waste less compute on the padding, which is faster for datatasets with significantly different system sizes or when a few structures are considerably larger than the median.
The bin sizes we have found to work well by default are 10 for the number of atoms and 2000 for the neighbor list.
We refer to the second option as ``per-batch-padded'' or ``pbp''. In each case, the preprocessing consists of shuffling the data, computing the neighbor list for all samples in a batch, applying padding, and stacking the input arrays.
We find that it is crucial for efficient \gls{mlip} training to compute the neighbor list as part of the preprocessing and not during the training step.
\gls{apax} uses the \code{vesin} library to compute neighborlists. 

In the ``cached'' version, a TensorFlow data pipeline is used to construct a \texttt{tf.data.Dataset} from a generator.
Here, all samples are processed in the first epoch, and ready samples are cached on disk using \texttt{tf.data}.
As a result, the data set is not saved in RAM, and batches are loaded asynchronously.
In the ``pbp'' version, we keep a buffer of prepared batches.
The buffer is asynchronously filled in a separate thread by a multiprocessing queue, which prepares batches on the fly, allowing the training step and batch preprocessing to overlap.

\subsection{Model Abstractions}
Separating an ML model implementation into distinct parts increases the modularity and extendability of a framework.
The distinct parts are referred to as `abstractions', whereby the central one in \gls{apax} is the \code{EnergyModel}. 
Internally, it consists of what we refer to as a representation, readout, and scale-shift layers.
To ease the training process, empirical energy correction terms, such as the Ziegler--Littman--Biersack potential\cite{zieglerStoppingRangeIons1985}, can be supplied.
A representation is any function from pairwise distances, atomic numbers, and the neighbor list to a per-atom feature vector.
In \gls{apax}, we currently implement the Gaussian Moment descriptor and an equivariant message passing model similar to NequIP\cite{batznerE3equivariantGraphNeural2022}, based on an example from the e3x\cite{unkeE3XEquivariantDeep2024} documentation.
The readout module consists of neural networks and computes per-atom predictions, such as atomic energies, from the previously obtained feature vectors.
Notably, shallow ensembling is implemented via the final readout module, and the \code{EnergyModel} does not sum over the ensemble axis.
The final atomic predictions are scaled and shifted by potentially element-dependent, learnable parameters.
Common choices for the initialization of these parameters are a least-squares regression\cite{zaverkinFastSampleEfficientInteratomic2021} or supplying fixed reference values for isolated atom energies\cite{batatiaDesignSpaceE3Equivariant2022}.

\begin{figure}
    \centering
    \includegraphics[width=1.\linewidth]{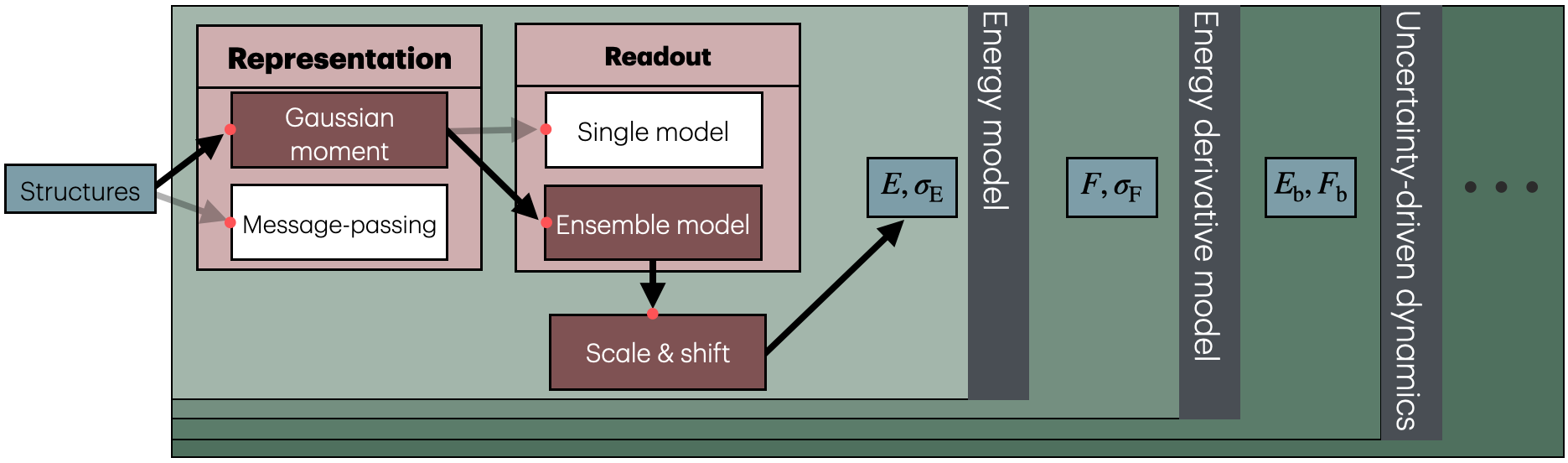}
    \caption{Schematic representation of the model abstractions and interaction with other functionalities in \gls{apax}. Inputs and model outputs added with each transformation are contained in the gray boxes.
    \label{fig:model_schema}}
\end{figure}

Decoupling the representation from distance calculations and energy predictions has two advantages:
First, it allows all functionalities in \gls{apax} to be independent of the chosen representation, allowing any new representation to instantly access all package capabilities.
Second, model developers can focus on the relevant parts that change between architectures.
The capabilities of an \code{EnergyModel} are then expanded by a series of function transformations.
For example, the gradient of the energy with respect to positions and the strain tensor allow the prediction of forces and virials, respectively.
The calculation of gradients is achieved by utilizing JAX's automatic differentiation capabilities.
Other features such as post hoc calibration, \gls{udd}, and the feature maps required by \gls{bal} are similarly implemented as function transformations.
The model abstractions discussed above are schematically represented in \cref{fig:model_schema}.

\subsection{Training}
In addition to model- and data pipeline-specific features, the trainer allows for flexible choices of optimizers, learning rates, and loss functions.
We interface with Optax\cite{deepmind2020jax}, a widely used library for stochastic optimization.
Optax implements optimizers, such as SGD\cite{Goodfellow-et-al-2016}, Adam\cite{DBLP:journals/corr/KingmaB14} and SAM\cite{foretSharpnessAwareMinimizationEfficiently2021}, and we have additionally added the recently proposed AdEMAmix\cite{pagliardiniAdEMAMixOptimizerBetter2024} algorithm.
A few Optax optimizers require special treatment in the training step and are not available.
In each case, they can be selected from the configuration file, with the option to supply optimizer-specific keyword arguments.
Further, \gls{apax} offers a linear decay and a cyclic cosine learning rate scheduler compatible with any optimizer.
Model parameters are grouped to allow separate learning rate selection for neural networks, embedding, and other parameters.
The loss factory provides a set of probabilistic and non-probabilistic loss functions, all of which can be used individually or linearly combined.
For each loss term, it is possible to set a weighting in the overall loss function and options to divide it by the number of atoms in a structure.

To accelerate training, \gls{apax} makes use of JAX's sharding API to automatically use all available GPUs or accelerator hardware, known as devices, for data parallelism.
For very large models, strategies such as pipeline\cite{huangGPipeEfficientTraining2019} and tensor parallelism\cite{shazeerMeshTensorFlowDeepLearning2018} are required.
These strategies are not implemented as speed requirements in \gls{md} simulations, essentially ruling out too large models as impractical.
During training, there are several options for tracking metrics.
\gls{apax} makes use of callbacks, which allow the logging of training metrics to CSV files, TensorBoard, and MLFlow. Further, when using IPSuite\cite{zillsCollaborationMachineLearnedPotentials2024}, metrics can additionally be monitored using Data Version Control.
For transfer learning tasks, \gls{apax} offers an interface to choose which parameters to load (\cref{lst:transfer}), which of these to freeze during fine-tuning, and which to re-initialize.

\subsection{Deployment}\label{sec:deployment}
Models trained in \gls{apax} can be applied to various tasks a force field might be used for, such as \gls{md} and geometry optimization.
For these purposes, we provide an interface to \gls{ase} and implement an \gls{md} engine, using the neighborlist and thermostats from JaxMD.
The \gls{ase} calculator provided by \gls{apax} can be used in Python scripts like any other calculator, but was extended with methods for batch evaluation of structures.
While the \gls{ase} interface is the most flexible, making it particularly useful for quickly setting up custom simulations, the internal \gls{md} engine achieves the performance required for production scales.
\gls{ase} is bottlenecked by blocking host-device transfers happening at every time step and the use of NumPy-based integrators.
Such bottlenecks are avoided in the JAX-based \gls{md} engine, as the entire simulation loop takes place on the GPU and is JIT compiled.
Further, the transfer of configurations for trajectory writing also happens asynchronously by using JAX's non-blocking host callbacks, leading to excellent device utilization.
Both engines use the neighbor list provided by JaxMD.
For periodic systems with very small cell sizes, the minimum image convention used by JaxMD breaks down.
Whenever the cell is small enough that multiple images of a neighbor are present in the local environment of an atom, the \gls{ase} calculator falls back to using the \code{vesin} neighbor list.

While the \gls{ase} readable formats, such as extended XYZ, are used commonly for distributing datasets and training atomistic machine learning models, they are not optimized for disk size, reading, or writing speed, and are usually fairly restrictive in the kinds of data that can be stored.
For example, it is common in MD simulations with \glspl{mlip} to calculate the model's uncertainty during the simulation.
Uncertainties computed on the fly can be used to terminate simulations\cite{zills2024zntrack,zillsMachineLearningdrivenInvestigation2024,schranCommitteeNeuralNetwork2020}, detect rare events\cite{vandermauseFlyActiveLearning2020a}, or compute uncertainties in ensemble averages\cite{imbalzanoUncertaintyEstimationMolecular2021}.
H5MD, by contrast, is a file format intended for high-performance \gls{md} simulations and can be written asynchronously. 
The ZnH5MD package allows for writing \gls{ase} Atoms objects to and reading them from H5MD files. In addition, all properties of the calculator results of the \gls{ase} Atoms objects can be stored. They are seamlessly (de)serialized with all other information in both the \gls{ase} and internal \gls{md} engine.

\Cref{eq:jacobian} has a particularly convenient implication for \gls{md} simulations:
Since it is possible to evaluate the forces of a model cheaply, without associated uncertainties (left-hand side) or more costly, with uncertainties (right-hand side), it is possible to switch between the two modes during a simulation.
In the JAX-based \gls{md} engine, we implement the switching such that the more expensive force uncertainty evaluation is only performed every $N$ steps, where $N$ is the dump interval of the trajectory.
As a consequence, well-calibrated force uncertainties can be calculated at almost no extra cost.

Finally, \gls{apax} comes with two possible ways of distributing models. First, model checkpoints can be uploaded to a cloud storage provider.
Second, tight integration with the ZnTrack\cite{zills2024zntrack} ecosystem allows for the straightforward download of models with a single line of code.
Models distributed \textit{via} ZnTrack can be used directly in ZnTrack/IPSuite\cite{zillsCollaborationMachineLearnedPotentials2024} workflows or standalone applications.
The models created for the present work are available via \verb|zntrack.from_rev(<model_name>, remote="https://github.com/apax-hub/apax_paper")|.

\section{Results and Discussion}

\subsection{Batch Data Selection}
\begin{figure}
 \centering
 \includegraphics[width=1.0\linewidth]{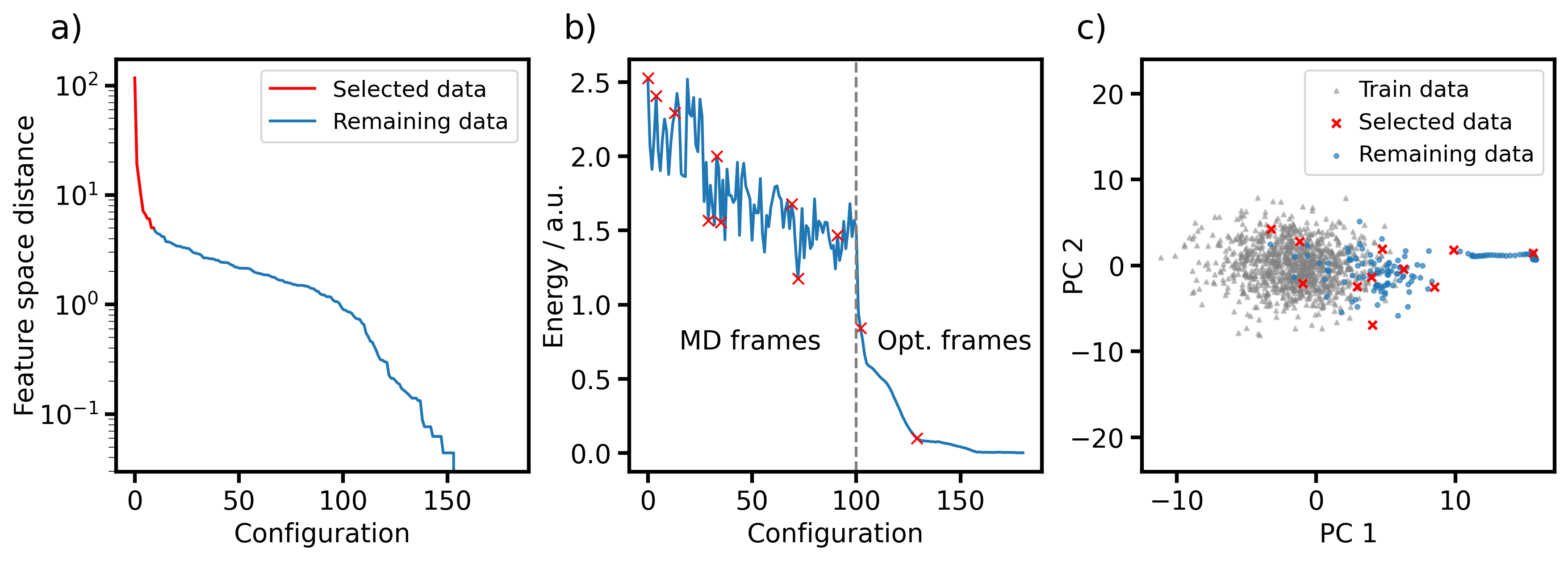}
 \caption{
    Selection heuristic and analysis of a combined \gls{md} and geometry optimization trajectory.
    a) sorted squared distances used in the MaxDist selection method. b) Energy for each configuration of the combined trajectory. c) First two principal components of the last-layer features for training data and data pool.
    The pool of data points is marked in blue, selected configurations are marked in red, and training configurations are marked in grey in each subplot.
 }
 \label{fig:bal}
\end{figure}

\gls{apax} augments the batch data selection methods it implements with visualizations to determine how many new data points should be selected.
To illustrate the selection process, we train a \gls{gmnn} model on a 1000-structure subset of the alanine tetra-peptide dataset from the MD22 collection\cite{chmielaAccurateGlobalMachine2023}. 
We use the resulting model to perform a \SI{10}{ps} \gls{md} simulation of that molecule, followed by a geometry optimization.
The data pool is subsequently constructed from the entire optimization trajectory and every 200th configuration from the \gls{md} trajectory.
10 samples are selected using maximum distance selection with last-layer features.
During the batch selection process, data points are ranked by the squared feature-space distance from every previously selected data point and all training samples.
By inspecting the sorted nearest-neighbor distances, it is possible to estimate how many highly informative data points the pool contains.
\cref{fig:bal}a) displays the feature distances for the \gls{md} trajectory generated above.

The combined trajectory for the data pool is shown in \cref{fig:bal}b) with the selected configurations highlighted as red dots.
Near convergence, geometry optimizations typically contain many very similar configurations, which the batch data selection method detects automatically.
As a result, most configurations are selected from the \gls{md} trajectory and only 2 configurations from the geometry optimization, none of them from the very end.
Further, the similarity of geometry optimization configurations is visible in the features used during the selection.
\cref{fig:bal}c) displays the first two principal components of the last-layer features for training and pool configurations, where the geometry optimization configurations are grouped closely together on the right-hand side.

\subsection{Continuous Learning}
The following section compares two active learning schemes using batch data selection, focusing on the ionic liquid \gls{emim}.
In the first approach, a new model is initialized and trained from scratch at each iteration. The second is a continuous learning approach in which we retrain the model starting from the parameters of the previous iteration, i.e., a \gls{tl} step is performed at each iteration. Although more advanced continuous learning techniques are available \cite{eckhoffLifelongMachineLearning2023a, quRecentAdvancesContinual2021}, this approach represents a robust baseline for such techniques.

Within four active learning workflows, \gls{gmnn} models are iteratively trained on data generated by the MACE-MP0 foundation model\cite{batatiaFoundationModelAtomistic2023}, which serves as the ground truth for this experiment.
The four workflows are initialized with the same core model, trained on the energies and forces of 40 structures, and validated on 20 structures sampled from a MACE-MP0 trajectory. Every structure consists of 480 atoms with periodic boundary conditions. Subsequently, the models of one workflow are retrained for 1000 epochs at each iteration, referred to as the R1000$_i$. The models of the other workflows underwent \gls{tl} for 150, 300, and 500 epochs to analyze the effect of the number of epochs in continuous learning approaches. In analogy, they are named CL150$_i$, CL300$_i$, and CL500$_i$. The number $i$ after each workflow label refers to the learning iteration.
Besides the number of epochs, all other hyper-parameters are identical across all models. The \gls{tl} models are trained on all available data. Further, no model parameters are frozen during \gls{tl}, as early tests revealed that the freezing of parameters degrades performance. More detailed setup is provided in the SI \cref*{SI:lotf}. Previous studies\cite{zaverkinTransferLearningChemically2023}, where \gls{tl} is applied to transfer learned features from a lower quality quantum chemistry method to a higher quality one, found that it is necessary to freeze most of the model weights, retraining only the last or last few layers, to prevent catastrophic forgetting. In contrast to our setup, the dataset size of the target level of theory is often small, and no extended dataset can be used, amplifying such problems.

\begin{figure}
   \includegraphics[width=0.5\linewidth]{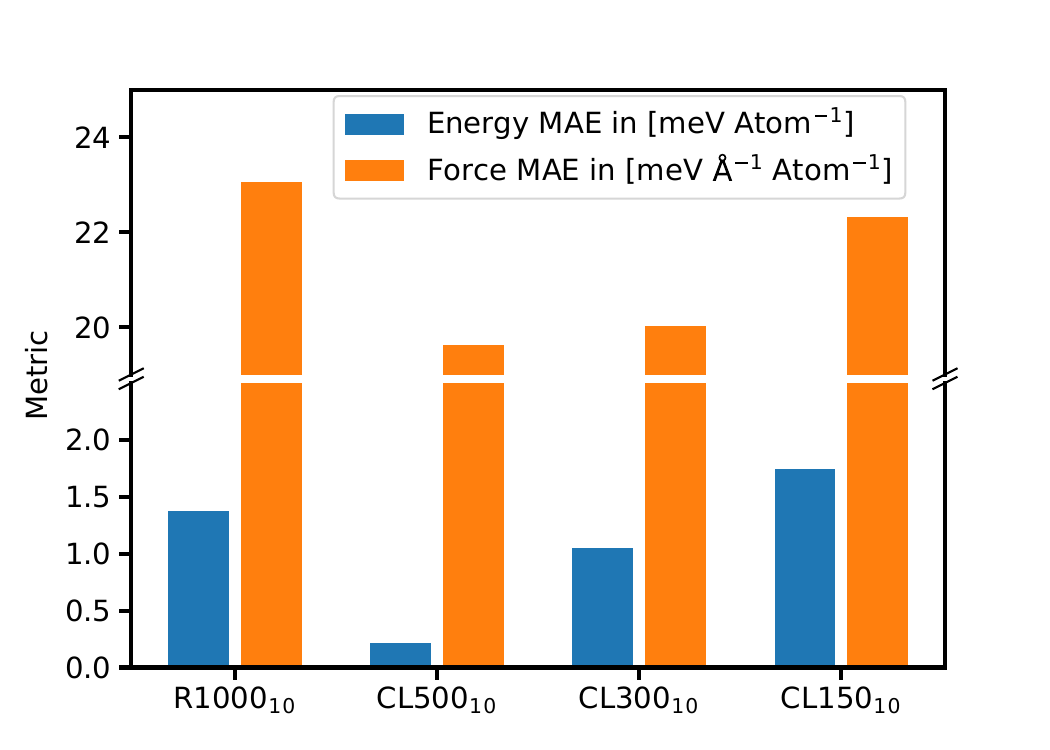}
   \caption{Evaluation metrics of CL150$_{10}$, CL300$_{10}$, CL500$_{10}$ and R1000$_{10}$ for a test set of the MACE-MP0 production trajectory.}
   \label{fig:method_vs_loss}
\end{figure}

For each of the four workflows, 240 training and 120 validation structures are accumulated by expanding the training dataset with 20 and the validation dataset with 10 structures over 10 learning iterations.
The structures are sampled from $\SI{200}{\pico\second}$ \gls{md} trajectories generated with the models from previous iterations.
Sampling simulations are performed with temperature profiles oscillating between $\SI{300}{\kelvin}$ and $\SI{600}{\kelvin}$ to achieve more diversity in the trajectories.
As shown in the supporting information \cref{si:fig:metric_conv}, the metrics of the models converge with the number of active learning iterations.
Comparing CL150$_{10}$, CL300$_{10}$, and CL500$_{10}$ on 150 test structures sampled from a MACE-MP0 trajectory in \cref{fig:method_vs_loss}, we found that the model's loss and accuracy metrics decrease with an increasing number of \gls{tl} epochs.
Notably, the force \gls{mae} of CL150$_{10}$ is already smaller than the force \gls{mae} of R1000$_{10}$ at 150 transfer learning epochs.
CL300$_{10}$ achieves cumulative savings of 7,000 epochs across 10 iterations while surpassing the accuracy of R1000$_{10}$ on both metrics.
Calibrated force uncertainties are monitored throughout simulations. Only CL300$_{1}$ stopped after $\SI{50}{\pico\second}$ due to exceeding an uncertainty threshold of $\SI{3.0}{\eV\per\angstrom}$.

\begin{figure}
   \includegraphics[width=1.0\linewidth]{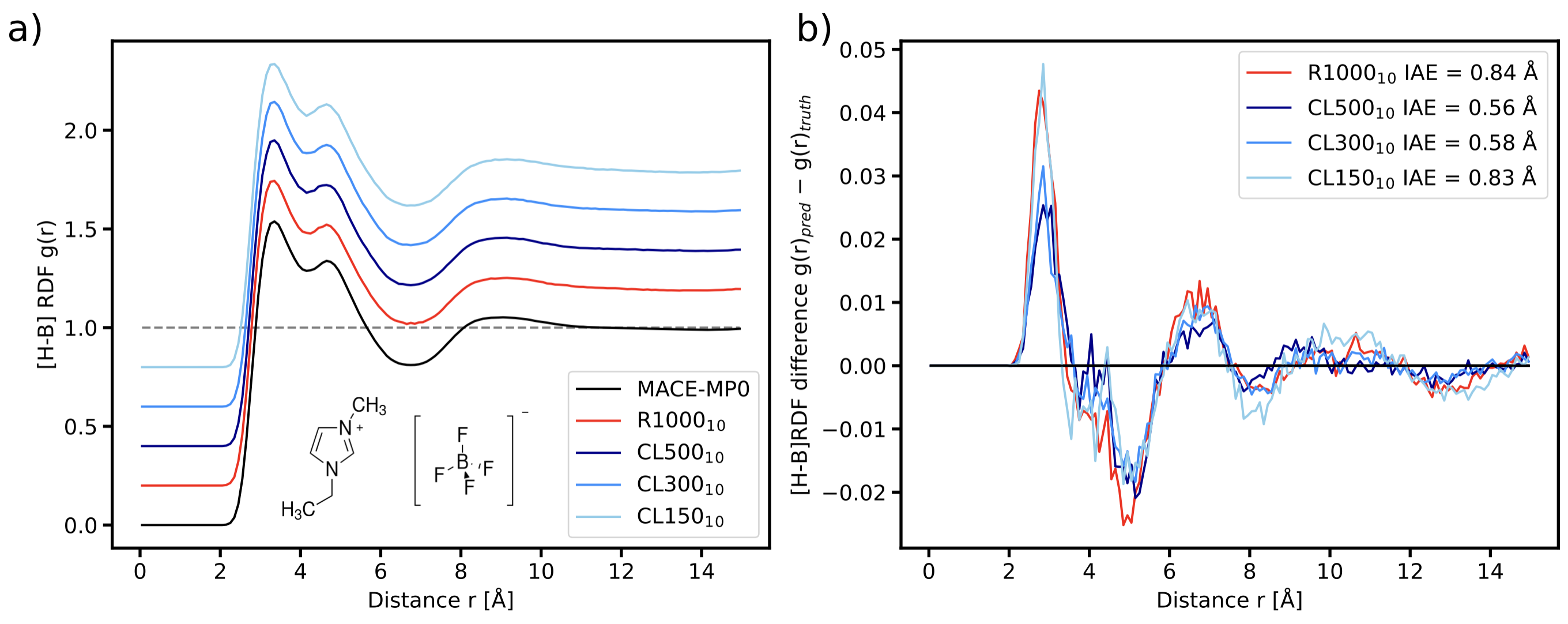}
   \caption{a) Hydrogen-boron \glspl{rdf} of \gls{emim} trajectories produced with the final models CL150$_{10}$, CL300$_{10}$, CL500$_{10}$ and R1000$_{10}$ and the MACE-MP0 foundation model serving as ground truth. For better visibility, the \glspl{rdf} are shifted with an offset of $0.2$. b) \glspl{rdf} differences of the final models and MACE-MP0 foundation model.}
   \label{fig:emim_rdf}
\end{figure}

Further, sampled quantities rather than model metrics are analyzed.
The inter-molecular hydrogen-boron \glspl{rdf} are plotted in \cref{fig:emim_rdf}a) for the final models CL150$_{10}$, CL300$_{10}$, CL500$_{10}$, and R1000$_{10}$. The \glspl{rdf} are based on $\SI{1}{\nano\second}$ simulations of an \gls{emim} system containing 2400 atoms at $\SI{400}{\kelvin}$ with time steps of $\SI{0.5}{\femto\second}$.
The difference between the reference \gls{rdf} and the ones simulated with the models CL150$_{10}$, CL300$_{10}$, CL500$_{10}$ and R1000$_{10}$ are displayed in \cref{fig:emim_rdf}b).

Even the least accurate of the four resulting models successfully reproduces the \glspl{rdf} of the reference potential with a maximal difference of around $\Delta g(r) = 0.05$. The only peaks that are shifted within the accuracy of the bin size of $\SI{0.1}{\angstrom}$ are the third flat peaks of CL500$_{10}$ and CL150$_{10}$, with a deviation of $\SI{-0.1}{\angstrom}$.
For further analysis, we compare the \gls{iae}, i.e., the absolute difference between the predicted RDF and the ground-truth RDF integrated over the distance up to $\SI{15}{\angstrom}$.
The \glspl{iae} of the H-B \glspl{rdf} decreases with an increasing number of transfer learning epochs, also reported in \cref{fig:emim_rdf}b).
However, that trend does not hold for some of the atom pairs' \glspl{rdf}, as shown in \cref{si:fig:more_emim_rdfs}.
The results demonstrate that the continuous learning approach can surpass the accuracy of retrained models for validation metrics and observables while saving considerable training time.

\subsection{Training Parallelization}
\begin{figure}
    \centering
    \includegraphics[width=0.5\linewidth]{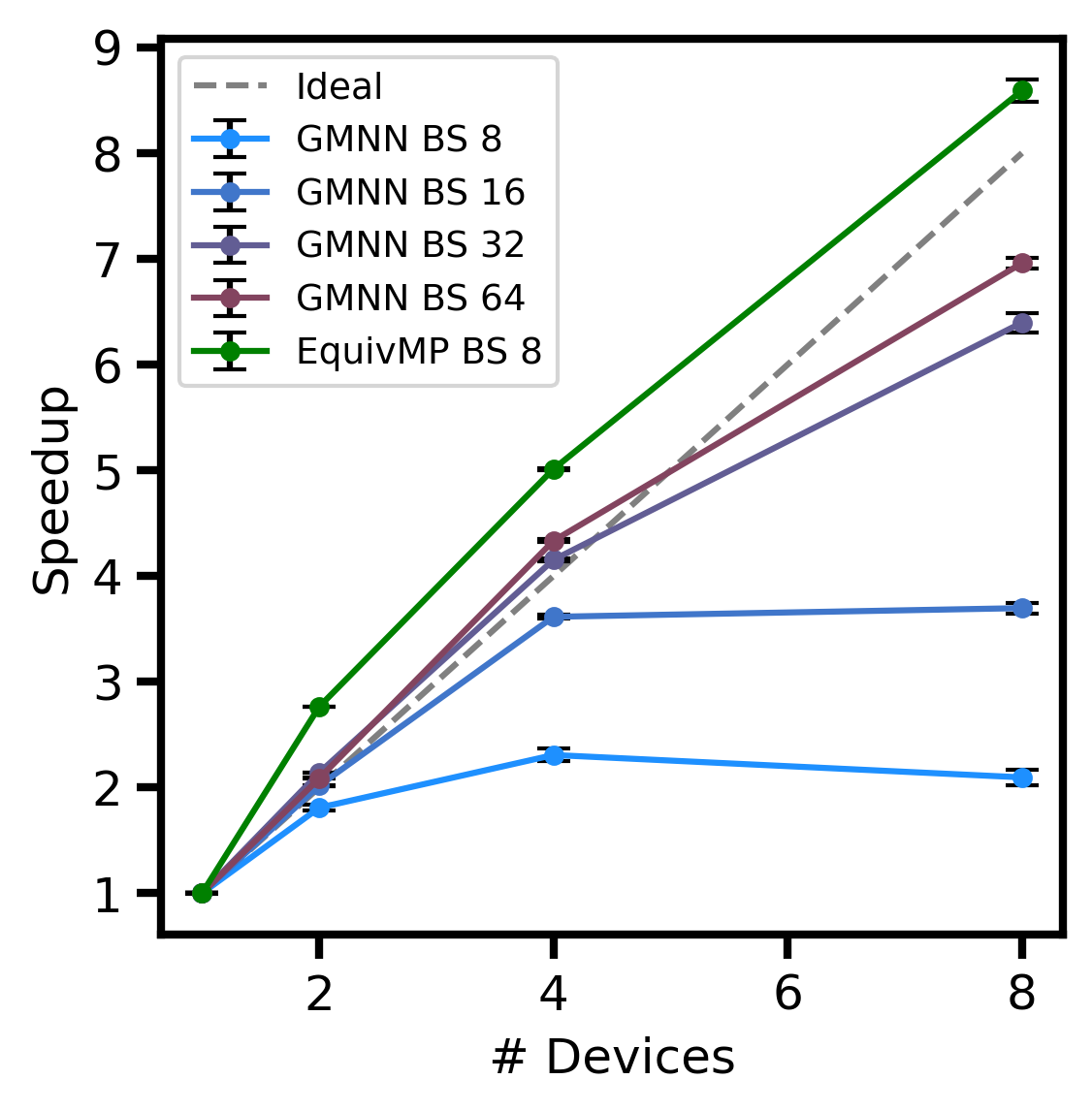}
    \caption{Speed-up of data parallel training compared to a single device for \gls{gmnn} and EquivMP models for batch sizes between 8 and 64 and 1, 2, 4, and 8 devices.}
    \label{fig:dptraining}
\end{figure}

We investigate the efficiency of the data parallel training strategy by training \gls{gmnn} and EquivMP models on the \gls{emim} dataset created in the previous section.
Training times are benchmarked with up to 8 GPUs and batch sizes between 8 and 64.
The selected range spans from 8 samples to 64, where the former is the minimum batch size capable of utilizing all 8 devices and the latter is the largest power of 2 that fits into the VRAM of a single device for the \gls{gmnn} model.
Both models used a 6.0~\AA{} radial cut-off and 7 radial basis functions. 
The \gls{gmnn} model used 5 features for the tensor contractions and two hidden layers with 256 units each.
The EquivMP model used a maximal rotation order of $L=2$, 32 channels, and 2 message-passing steps.
All runs are performed on an NVIDIA DGX node with 8 A100 accelerators running CUDA 12.2 and jaxlib 0.4.35.
The speed-up in time per epoch compared to training the same model at the same batch size on a single device is displayed in \cref{fig:dptraining}.
The epoch times are averaged over a training run of 20 epochs, removing the first from the average as it contains the initial compilation time, which is mostly independent of the number of devices.
The error bars are the standard error of the mean calculated from these timings.

At a batch size of 8, the \gls{gmnn} model achieves 90 \% of the ideal speed-up on 2 devices.
As the number of devices increases at a constant compute load, the communication overhead increases.
Since \gls{gmnn} is an extremely lightweight model, training does not saturate the A100s at low sample sizes per device and the communication overhead is significant, which we also observe for the batch size 16 GMNN model on 8 devices.
However, at the larger batch sizes 32 and 64, the scaling remains well across all device counts, with its lowest value being 75 \% of the ideal speed-up.
In the case of training on 4 devices, we observe a better-than-ideal scaling at these batch sizes.
The same behavior can be observed for the EquivMP model for all device counts larger than one.
We attribute better-than-ideal scaling behavior to the generation of more efficient GPU kernels by the XLA compiler at smaller sample sizes per device.
It should be noted that the expected speedup depends on the system sizes used in the training set and the size of the model. Naturally, for smaller systems and models, the speedup will be smaller than what is reported here.

\subsection{Inference Performance}
\begin{figure}
    \centering
    \includegraphics[width=1.0\linewidth]{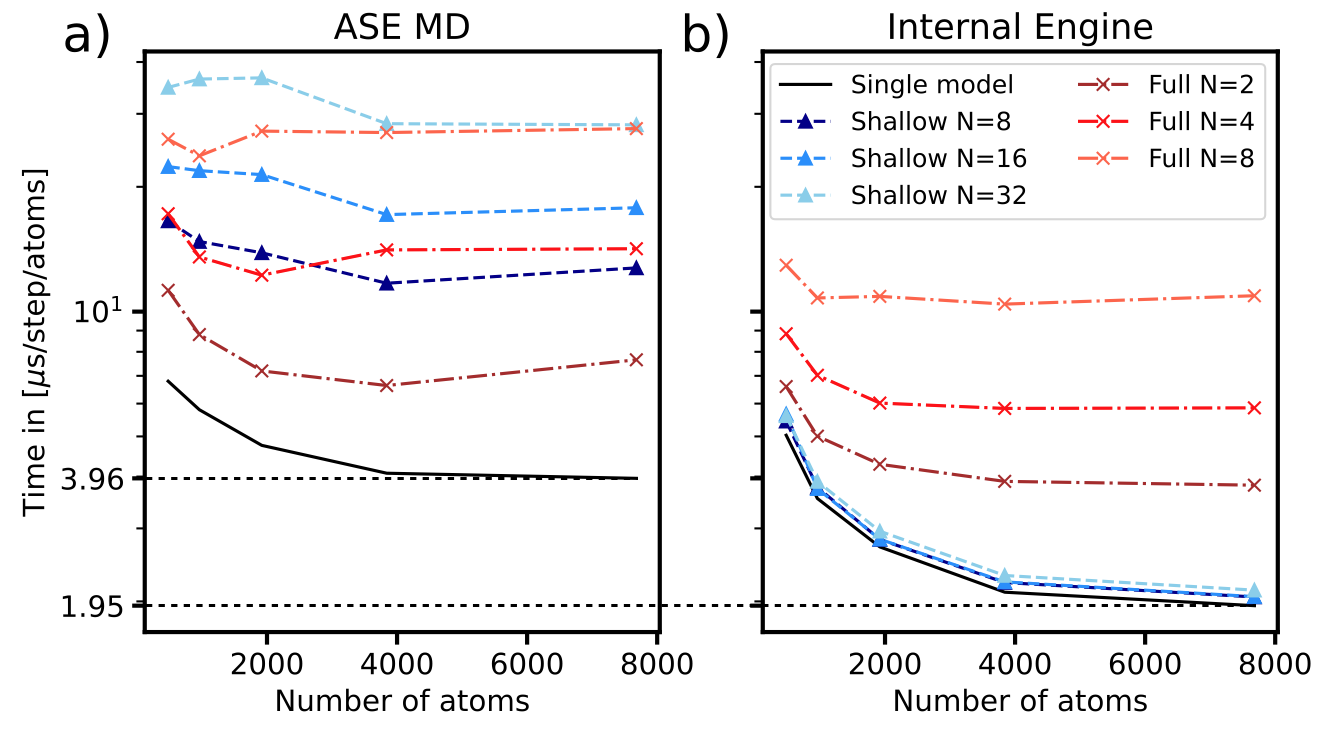}
    \caption{Inference performance of full and shallow ensemble models measure in \si{\micro\second\per\atom\per\step} for various system sizes. a) timings for the \gls{ase} calculator. b) timings for the internal \gls{md} engine.}
    \label{fig:perfomance}
\end{figure}

Various ensemble models, along with a single model consisting of one set of parameters, are trained on \gls{emim} data to evaluate the inference performance of the JAX-based \gls{md} engine and \gls{ase}. Additionally, a comparative analysis is conducted between the two available ensemble methods: full and shallow.
For each model, five inference times corresponding to different system sizes (480, 960, 1920, 3840, and 7680 atoms) are illustrated in \cref{fig:perfomance}. These times are averaged over 50,000 \gls{md} steps simulated on an RTX 4090 with a step size of \SI{0.5}{\femto\second}.

The first kind of model is the full ensemble.
For both \gls{md} engines, using ensembles with $N_{\text{ens}}$ members is more efficient than $N_{\text{ens}}$ separate models across all system sizes.
Doubling $N_{\text{ens}}$ does not result in twice the inference time, instead, it can reduce the step-time per ensemble member by up to 30\%, depending on the size of the system and type of ensemble.
The best inference time per member reduction compared to a single model is 78\% for 480 atoms and $N_{\text{ens}}$=8 members while the least reduction is 7\% for $N_{\text{ens}}$=2, resulting in near linear scaling for the largest structure with 7680 atoms.

While the costs of the thermostat and neighbor list updates are constant across ensemble sizes for each system, they are negligible compared to the evaluation of even a single model.
Thus, the observed performance gains can be attributed to the generation of more efficient CUDA kernels by the \gls{xla} compiler.

The shallow ensemble models with $N_{\text{ens}}$ members outperform $N_{\text{ens}}$ separate models considerably.
The smallest step-time per ensemble member reduction is $60\%$ compared to a single model.
As shown in \cref{fig:perfomance}a), with the \gls{ase} \gls{md} engine, they have smaller inference times compared to full ensembles with the same number of ensemble members.
With the internal \gls{md} engine, see \cref{fig:perfomance}b), shallow ensembles can make use of the switching of evaluation modes introduced in \cref{eq:jacobian}.
Thus, their prediction times are nearly independent of the number of ensemble members $N_{\text{ens}}$, and are, with minor deviations, as fast as a single model.
At large system and ensemble sizes, the GPU is saturated, and the simulation time scales linearly with the number of atoms. 
For smaller sizes, the scaling is sub-linear.
Overall, the internal \gls{md} engine consistently outperforms \gls{ase} in inference performance.
The performance advantage arises primarily due to the JIT compilation of the simulation loop, which unlike \gls{ase} is computed entirely on the GPU and the optimizations described in \cref{sec:deployment}.
Performance inconsistencies in \gls{ase} inference times are attributed to unoptimized memory transfer processes.

Facilitating comparisons between models and implementations on equal grounds is challenging as performance characteristics vary with the number density of different systems and the hardware used to conduct the simulations.

Hence, we consider the example of the \ce{Li3PO4} solid-state electrolyte\cite{musaelianLearningLocalEquivariant2023}, for which evaluation metrics and \gls{md} simulation speeds of Allegro\cite{musaelianLearningLocalEquivariant2023} and SO3krates\cite{frankEuclideanTransformerFast2024} were reported using the same GPU, an Nvidia V100.
We follow the training setup of \citeauthor{musaelianLearningLocalEquivariant2023}\cite{musaelianLearningLocalEquivariant2023} and report energy and force test \glspl{mae} as well as the time per \gls{md} step in \cref{tab:li3po4_perf}.
The \gls{md} simulation lasted \SI{50}{\pico\second} and is performed at \SI{600}{\kelvin}.
It should be noted that the timings include the entire \gls{md} step, not just the model evaluation.
Allegro timings were measured in LAMMPS, So3Krates in \code{mlff}\cite{frank2022so3krates} and GMNN in \gls{apax}' internal \gls{md} engine.

\begin{table}[H]
  \caption{Test errors and inference times of Allegro, SO3krates and \gls{gmnn} models trained on a 10K subset of the \ce{Li3PO4} dataset. The inference time is reported as the wall time per time step per atom for a 192 and a 5154 atom system. All simulations are performed on a Nvidia V100.
  }
  \label{tab:li3po4_perf}
  \begin{tabular}{lrrrr}
    \toprule
     & E \gls{mae} / \si{\milli\electronvolt\per\atom} & F \gls{mae} / \si{\milli\electronvolt\per\angstrom}  & \#atoms &  \si{\micro\second\per\atom\per\step}  \\
    \midrule
    % Allegro     &  1.7 & 73.4 & 192  & 27.8 \\
    Allegro     &  1.7 & 73.4 & 192  & 27.8 \\
    SO3krates   &  0.2 & 28.2 & 192  & 23.6 \\
    GMNN &  0.8 & 68.9 & 192  & 10.1 \\
    GMNN &  0.8 & 68.9 & 5184 & 2.3 \\
    \bottomrule
  \end{tabular}
\end{table}

In terms of accuracy, \gls{gmnn} falls between the Allegro and SO3krates models.
However, the simulation speed for the 192-atom system is more than 2 times higher than SO3krates  and almost 3 times higher than Allegro.
It should be noted that in the case of \gls{gmnn}, the GPU is not saturated for 192 atoms.
We repeat the simulation for a $3\times 3\times 3$ supercell (5184 atoms) where \gls{apax} reaches \SI{2.3}{\micro\second\per\atom\per\step}, outperforming the other models by more than a factor of 10.
The resulting \gls{rdf} is reported in \cref{si:fig:li3po4-rdf} and shows good agreement with the \gls{dft} reference.
A common consideration in computational materials science is the accuracy-speed trade-off of various simulation methods.
At least for \ce{Li3PO4} and the observable considered here, all models are in good agreement with the \gls{dft} reference.
Thus, additional accuracy beyond the Allegro model is not required, while more performant models can significantly reduce the time to solution.

\subsection{Uncertainty-Driven Dynamics}
\begin{figure}
    \centering
    \includegraphics[width=0.5\linewidth]{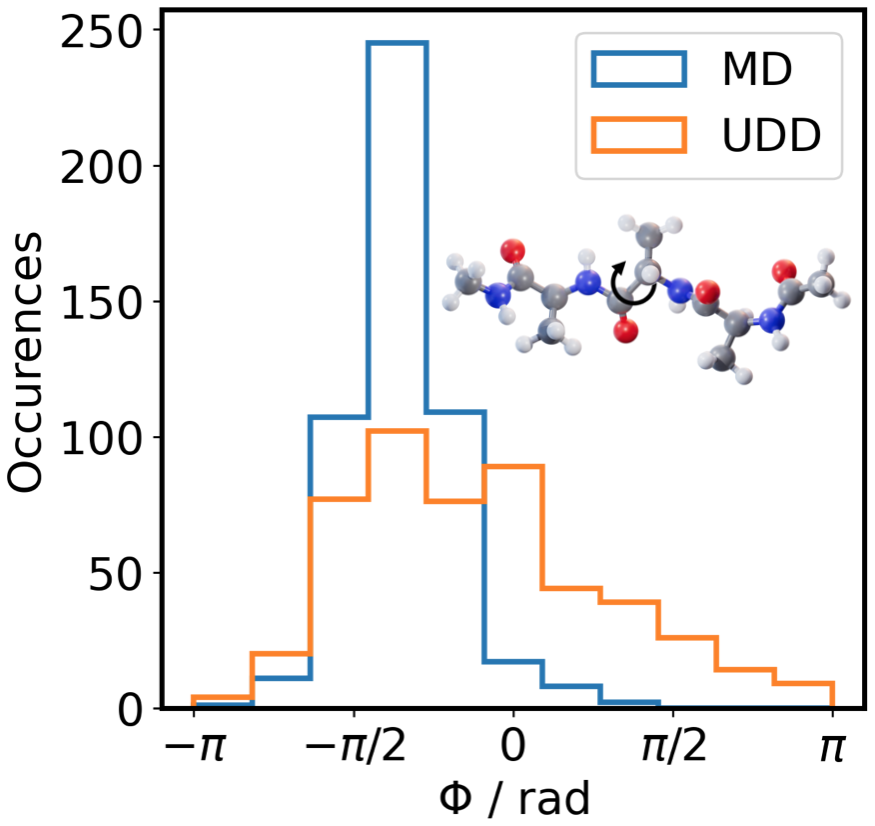}
    \caption{Comparison of alanine tetrapeptide dihedral angle distributions from short unbiased \gls{md} and \gls{udd} trajectories.}
    \label{fig:udd}
\end{figure}

Finally, we illustrate the flexibility of \gls{apax} by training an equivariant message passing model as a shallow ensemble and use it to perform \gls{udd}.
We train an ensemble of 8 shallow members on the alanine tetrapeptide dataset from the MD22 collection\cite{chmielaAccurateGlobalMachine2023}.
Two trajectories are simulated at \SI{300}{\kelvin} using a Langevin thermostat for \SI{50}{\pico\second}.
The first one is an unbiased \gls{md} run, while the latter uses a \gls{udd} bias following \cref{eq:udd} with $A = \SI{1}{\eV\per\atom}$ and $B = \SI{1.2}{\eV}$.
To analyze the effect of the enhanced sampling method, we consider the distribution of one dihedral angle during the trajectories.
The atoms involved in the dihedral angle and the corresponding histogram are displayed in \cref{fig:udd}.

In the unbiased \gls{md} run, the dihedral spends most of the time near the energy minimum it started from.
The biased trajectory, on the other hand, explores the full rotation around the dihedral angle and can flatten out the histogram significantly, even after such a short simulation.

\section{Conclusion}
    
In this work, we presented \gls{apax}, a flexible, easy-to-use, and high-performance framework for training and deploying \glspl{mlip} with a focus on active learning. Beyond showcasing its capabilities, it delivers superior performance with streamlined functionality, essential for in silico experiments and rapid method development.
It is based on the JAX numerical computing library, which allows it to run on GPUs and utilize automatic differentiation. The model abstractions implemented in \gls{apax} allow for the straightforward integration of existing JAX-based architectures. Once integrated, they automatically have access to all features in the package, such as batch data selection methods, model ensembling, and \gls{udd}. The extendability of \gls{apax} is achieved through a combination of the powerful function transformations provided by JAX, which are extended by our own.

We have demonstrated the efficacy of \gls{apax} in active-learning scenarios by highlighting several aspects of the package.
First, we show how a continuous learning approach can be used to increase model accuracy while simultaneously lowering the number of training epochs compared to training from scratch.
Second, in data-parallel settings, \gls{apax} achieves great scaling with the number of devices for \gls{gmnn} at larger batch sizes and for the equivariant message passing model even at lower ones.
For inference, the \gls{gmnn} model achieves better accuracy than a performance-optimized Allegro while still being between 3 and 12 times faster.
Further, when deploying shallow ensembles, the cost of force uncertainty estimation can be made negligible by switching between evaluation modes.
By only evaluating force uncertainties on collected configurations, the inference speed of shallow ensembles is essentially identical to a single model.
As shallow ensemble uncertainty estimates converge quickly with ensemble size, they only incur a minor training time increase compared to traditional, full ensembles.
Finally, we illustrated the flexibility of \gls{apax}'s model abstractions by training an equivariant message-passing model as a shallow ensemble.
The model is composed with \gls{udd}, which was straightforward to implement as a model-agnostic function transformation.
The composed model is used to enhance the configurational sampling of alanine tetrapeptide.
In short, \gls{apax} is fast, extandable and well suited for active learning scenarios. It integrates tools like enhanced sampling techniques with selection methods in a straightforward manner.
Consequently \gls{apax} is well-suited for studying processes involving slow degrees of freedom, such as slow diffusion of liquids or separated metastable states of biomolecules.

The \gls{apax} package has been successfully used in the study of ionic liquids\cite{zillsMachineLearningdrivenInvestigation2024}, organic solvents\cite{zillsCollaborationMachineLearnedPotentials2024}, and \ce{CO2} hydrogenation \cite{LuukKempen2025CO2hydrogenation} with more applications currently being worked on.
Nonetheless, there are still opportunities for improvement.
While the conditional force uncertainty evaluation of shallow ensembles in the internal \gls{md} engine reduces inference time, it does not reduce the memory requirements, as a potentially large ensemble needs to be evaluated in memory.
The required memory could be reduced using techniques such as gradient checkpointing or evaluating the model in a loop instead of a vectorized manner.
A current limitation is the lack of multi-GPU support for \gls{md}.
As a result, the maximal system size that can be simulated is limited by the available VRAM on a single GPU. A possible solution is interfacing established MD codes such as Lammps\cite{thompsonLAMMPSFlexibleSimulation2022} via chemtrain-deploy\cite{fuchs_chemtrain-deploy_2025} or TinkerHP\cite{lagardereTinkerHPMassivelyParallel2018} via Deep-HP\cite{jaffrelot_inizan_scalable_2023}. \gls{apax} is under active development, with these and other additions in progress.

\section*{Associated Content}
\subsection*{Data and Software Availability}
The scripts and workflows needed to reproduce the work presented here can be found at \url{https://github.com/apax-hub/apax_paper}.
All data generated during the iterative training and production simulations are stored on an S3-object storage.
It can be obtained by cloning the git repository and executing \texttt{dvc pull} in the repository folder.
Further, the data can also be accessed on DaRUS \url{https://doi.org/10.18419/DARUS-5007}.

All software used throughout this work is publicly available.
The \gls{apax} repository is available on Github at \url{https://github.com/apax-hub/apax}  and can be installed from PyPi \textit{via} \texttt{pip install apax}.
IPSuite is available at \url{https://github.com/zincware/IPSuite} and can similarly be installed \textit{via} \texttt{pip install ipsuite}.

\subsection*{Supporting Information}
Configuration file example, validation error output example, training config of active learning experiments, model convergence metrics, \gls{emim} \glspl{rdf}, \ce{Li3PO4} \gls{rdf} (PDF)

\section*{Auther Information}
\subsection*{Corresponding Author}
\textbf{Johannes K\"{a}stner} - Institute for Theoretical Chemistry, University of Stuttgart, Pfaffenwaldring 55, 70569 Stuttgart, Germany; https://orcid.org/0000-0001-6178-7669; Email: kaestner@theochem.uni-stuttgart.de

\subsection*{Authors}

\textbf{Moritz R. Sch\"{a}fer} - Institute for Theoretical Chemistry, University of Stuttgart, Pfaffenwaldring 55, 70569 Stuttgart, Germany; https://orcid.org/0000-0001-8474-5808

\noindent\textbf{Nico Segreto} - Institute for Theoretical Chemistry, University of Stuttgart, Pfaffenwaldring 55, 70569 Stuttgart, Germany; https://orcid.org/0000-0003-3546-4879

\noindent\textbf{Fabian Zills} - Institute for Computational Physics, University of Stuttgart, Allmandring 3, 70569 Stuttgart, Germany; https://orcid.org/0000-0002-6936-4692

\noindent\textbf{Christian Holm} - Institute for Computational Physics, University of Stuttgart, Allmandring 3, 70569 Stuttgart, Germany; https://orcid.org/0000-0003-2739-310X

\subsection*{Author Contributions}
\textbf{M. R. Schäfer:} Conceptualization, Software, Investigation, Visualization, Data Curation, Writing - Original Draft, Writing – Review \& Editing

\noindent\textbf{N. Segreto:} Conceptualization, Software, Investigation, Visualization, Data Curation, Writing - Original Draft, Writing – Review \& Editing

\noindent\textbf{F. Zills:} Software, Validation, Data Curation, Writing – Review \& Editing

\noindent\textbf{C. Holm:} Resources, Writing – Review \& Editing, Supervision, Funding acquisition

\noindent\textbf{J. Kästner:} Resources, Writing – Review \& Editing, Supervision, Funding acquisition

\noindent M. R. Schäfer and N. Segreto contributed equally and share first authorship.

\subsection*{Notes}

The authors declare no competing financial interests.

\section*{Acknowledgements}

The authors would like to thank Lisa Schröder for insightful feedback on an early version of the manuscript.

C.H., J.K., F.Z., and M.S. acknowledge support by the Deutsche Forschungsgemeinschaft (DFG, German Research Foundation) in the framework of the priority program SPP 2363, “Utilization and Development of Machine Learning for Molecular Applications - Molecular Machine Learning” Project No. 497249646.

N.S. acknowledges support by the Deutsche Forschungsgemeinschaft project number 516238647 - SFB1667/1 (ATLAS - Advancing Technologies for Low-Altitude Satellites), and by the Ministry of Science, Research and the Arts Baden-Württemberg in the Artificial Intelligence Software Academy (AISA).

Further funding through the DFG under Germany's Excellence Strategy - EXC 2075 - 390740016 and the Stuttgart Center for Simulation Science (SimTech) was provided.

All authors acknowledge support by the state of Baden-Württemberg through bwHPC
and the German Research Foundation (DFG) through grant INST 35/1597-1 FUGG.
\setcounter{table}{0}
\setcounter{figure}{0}
\setcounter{equation}{0}
\setcounter{section}{0}

\renewcommand*{\thepage}{S\arabic{page}}
\renewcommand{\theequation}{S\arabic{equation}}
\renewcommand{\thetable}{S\arabic{table}}
\renewcommand{\thefigure}{S\arabic{figure}}
\renewcommand{\thesection}{S\arabic{section}}
\renewcommand{\thelstlisting}{S\arabic{lstlisting}}

\section{Configuration file and validation errors \label{SI:yaml}}

\Cref{lst:transfer} and \cref{lst:mdconfig} below give examples of the configuration files used for model training and inference using the internal \gls{md} engine in \gls{apax}, respectively.

\begin{lstlisting}[language=yaml, caption={Sections of an input file relevant for transfer learning  demonstrating the loading, freezing and re-initializing of parameters from a previously trained model.}, label={lst:transfer}]
checkpoints:
  # load existing model
  base_model_checkpoint: path/pre_trained_model
  # reinitialize scaling and shifting parameters
  reset_layers: [scale_shift] 

optimizer:
  emb_lr: 0.0000 # freeze embedding layer
  # train other parameters
  nn_lr: 0.0001  
  scale_lr: 0.001
  shift_lr: 0.0001
\end{lstlisting}

\newpage
\begin{lstlisting}[language=yaml, caption={Input file for performing a JAX-based MD simulation with an \gls{apax} model}, label={lst:mdconfig}]
ensemble:
  name: nvt
  dt: 0.5 # fs time step
  temperature_schedule:
    name: constant
    T0: <T> # K
  thermostat_chain:
    chain_length: 3
    chain_steps: 2
    sy_steps: 3
    tau: 100

duration: <DURATION> # fs
n_inner: 500 # compiled innner steps
sampling_rate: 10  # dump interval
buffer_size: 100
dr_threshold: 0.5 # Neighborlist skin
extra_capacity: 0

sim_dir: md
initial_structure: <INITIAL_STRUCTURE>
load_momenta: false
traj_name: md.h5
restart: true
checkpoint_interval: 50_000
disable_pbar: false

\end{lstlisting}

\gls{apax}'s command line interface has a built-in validator for its configuration files.
The validator is based on Pydantic and can catch missing or misspelled keywords and the usage of incorrect data types.
\Cref{lst:pydantic_val_error} shows the error messages when catching common mistakes in a training configuration file.

\newpage
\begin{lstlisting}[language=yaml, caption={Validation errors for a training configuration file in which the number of epochs keyword was misspelled and the training directory was not specified.}, label={lst:pydantic_val_error}]
apax validate train train.yaml
>>> 3 validation errors for config
>>> n_epochs
>>>   Field required
>>>   input_type: int
>>> nepochs
>>>   Extra inputs are not permitted
>>>   input_type: int
>>>   input: 1000
>>> data.directory
>>>   Field required
>>>   input_type: str
>>> Configuration Invalid!
\end{lstlisting}

\newpage
\section{Continuous Learning \label{SI:lotf}}

 In \cref{lst:h-params-rdf}, non-default hyper-parameters used in the active learning experiments are listed.
 A full model configuration with all default parameters can be found under\\
 \href{https://apax.readthedocs.io/en/latest/configs/full_configs.html}{https://apax.readthedocs.io/en/latest/configs/full\_configs.html}, state Nov. 28, 2024.
     
\begin{lstlisting}[language=yaml, caption={Non default hyper-parameter of the models used for the active learning experiments.}, label={lst:h-params-rdf}]
n_epochs: 1000 / 500 / 300 / 150

data:
  batch_size: 1
  valid_batch_size: 20

model:
  ensemble:
    kind: shallow
    n_members: 16

loss:
- name: energy
  loss_type: crps
- name: forces
  loss_type: crps

optimizer:
  name: adam
  emb_lr: 0.0005
  nn_lr: 0.0005
  scale_lr: 0.0005
  shift_lr: 0.0005
  schedule:
    decay_factor: 0.95

\end{lstlisting}

Here, we present the \gls{mae} and \gls{rmse} for both energy and force on a validation dataset, plotted against the learning cycles for all four active learning approaches.

     \begin{figure}[H]
        \centering
        \includegraphics[width=1.0\linewidth]{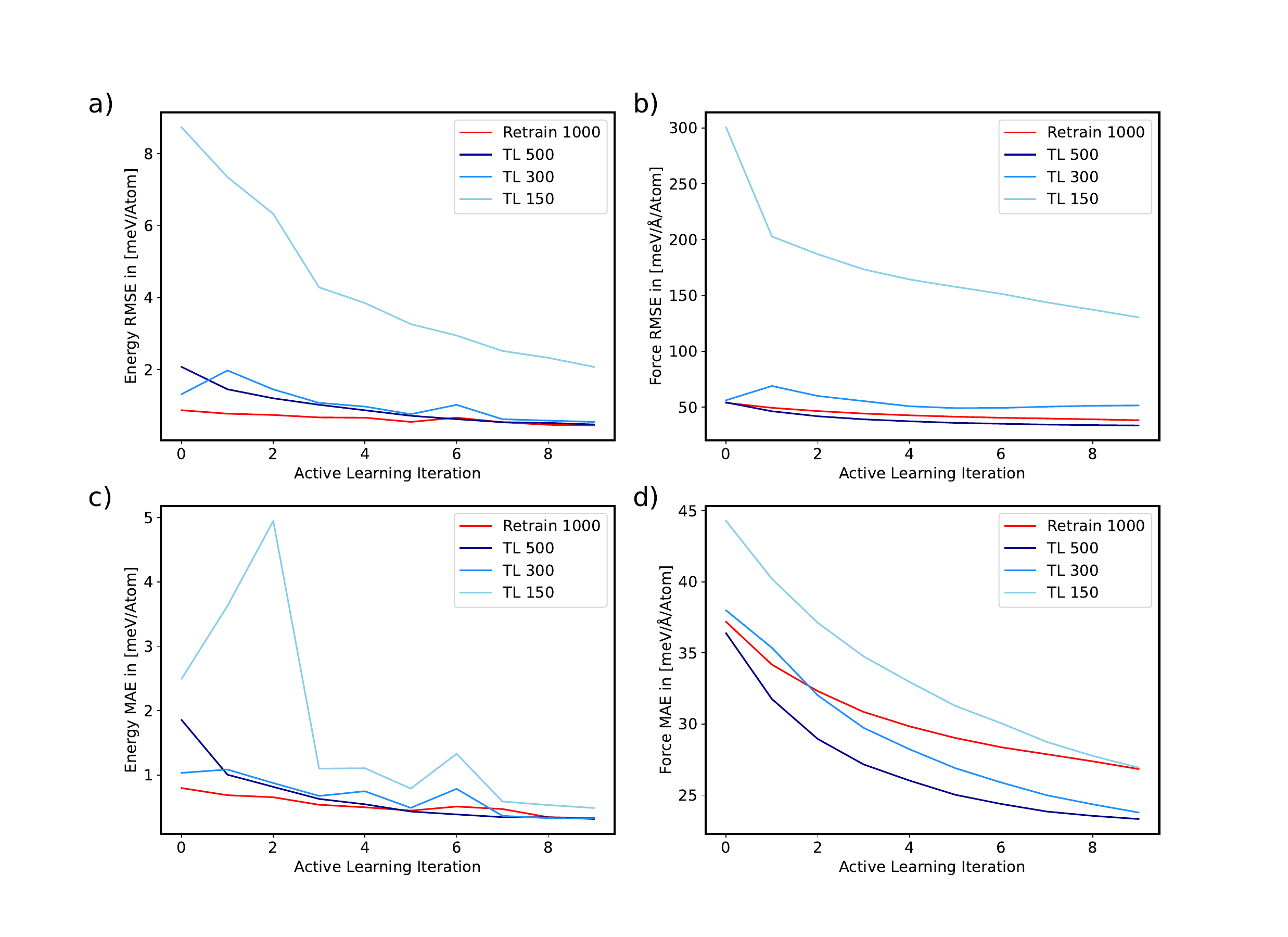}
        \caption{Validation metrics of all 4 active learning approaches plotted against their learning cycles. a) Energy \gls{rmse}, b) force \gls{rmse}, c) energy \gls{mae}, and d) force \gls{mae}.}
        \label{si:fig:metric_conv}
    \end{figure}

\begin{figure}[H]
        \centering
        \includegraphics[width=1\linewidth]{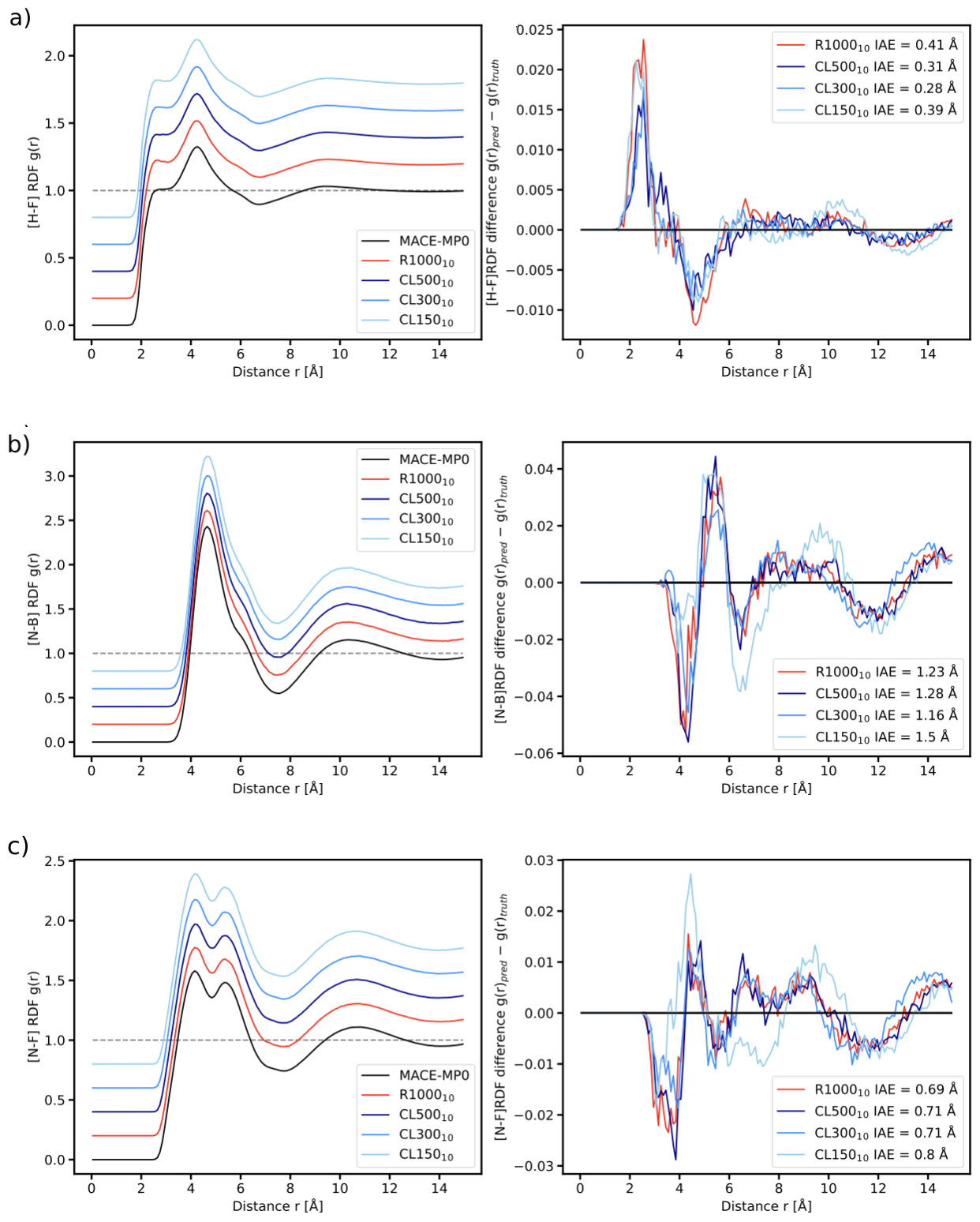}
        \caption{Radial distribution function of \gls{emim} at \SI{400}{\kelvin} obtained from \SI{1}{\nano\second} trajectories simulated with \glspl{mlip} compared to a MACE-MP0 reference.}
        \label{si:fig:more_emim_rdfs}
    \end{figure}

Furthermore, for completeness, all radial distribution functions (RDFs) of \gls{emim} and their corresponding errors are provided in \cref{si:fig:more_emim_rdfs}. These are based on foundational trajectories generated using the final models from the four active learning cycles, with the MACE-MP0 model serving as the reference.

 \section{Li\textsubscript{3}PO\textsubscript{4} Radial Distribution Function}

\begin{figure}
        \centering
        \includegraphics[width=0.6\linewidth]{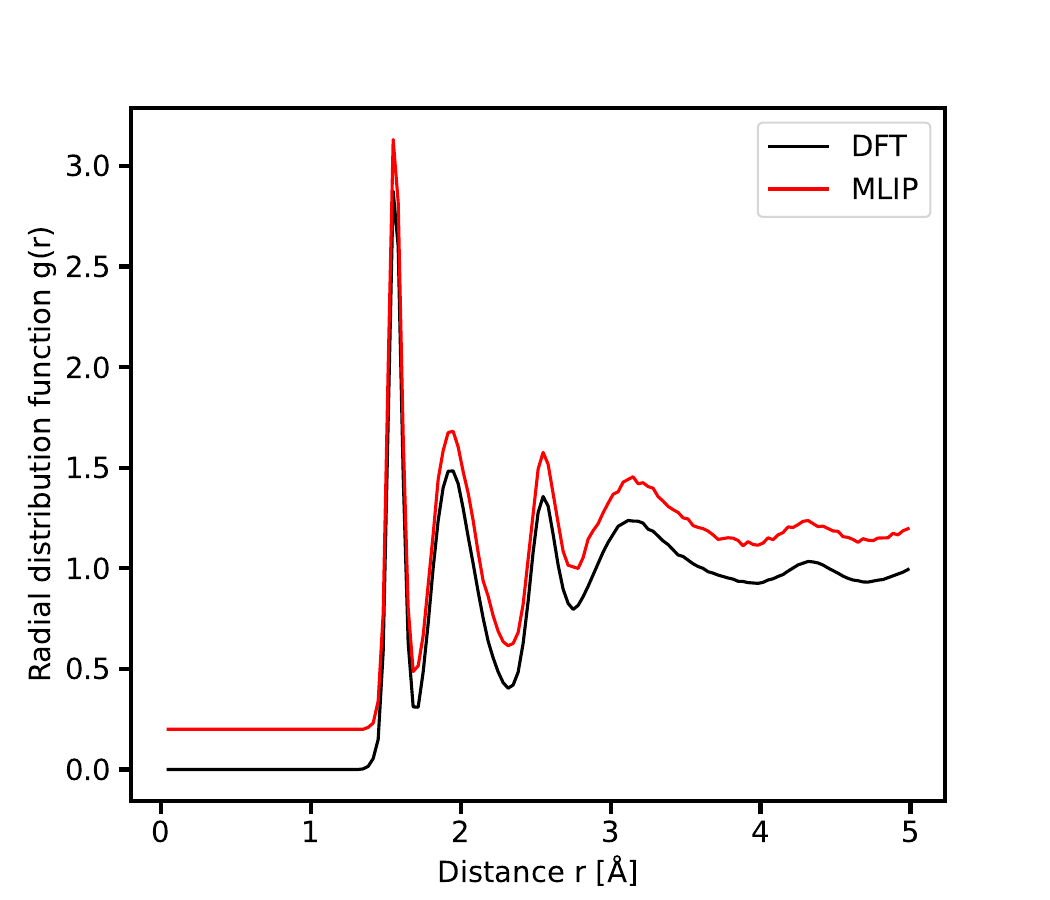}
        \caption{Radial distribution function of \ce{Li3PO4} at \SI{600}{\kelvin} obtained from a \SI{50}{\pico\second} trajectory simulated with a \gls{mlip} compared to a \gls{dft} reference. For visualisation, the \gls{mlip} \gls{rdf} is shifted about $0.2$.}
        \label{si:fig:li3po4-rdf}
    \end{figure}

    \Cref{si:fig:li3po4-rdf} displays the radial distribution functions obtained from the MD simulations of \ce{Li3PO4} using a GMNN model and a DFT reference.
    The DFT reference was computed from the \SI{600}{K} subset of the \ce{Li3PO4} dataset\cite{musaelianLearningLocalEquivariant2023}.
    The GMNN model agrees with the reference observable.
    
\bibliography{references}

\end{document}